\documentclass[twocolumn,english,nobibnotes,nofootinbib]{revtex4-1}
\usepackage[T1]{fontenc}
\usepackage[a4paper]{geometry}
\usepackage{float}
\usepackage{textcomp}
\usepackage{amsmath}
\usepackage{amssymb}
\usepackage{graphicx}
\usepackage{esint}
\usepackage{natbib}
\usepackage{color}

\makeatletter
\@ifundefined{textcolor}{}
{%
 \definecolor{BLACK}{gray}{0}
 \definecolor{WHITE}{gray}{1}
 \definecolor{RED}{rgb}{1,0,0}
 \definecolor{GREEN}{rgb}{0,1,0}
 \definecolor{BLUE}{rgb}{0,0,1}
 \definecolor{CYAN}{cmyk}{1,0,0,0}
 \definecolor{MAGENTA}{cmyk}{0,1,0,0}
 \definecolor{YELLOW}{cmyk}{0,0,1,0}
}

\begin{document}

\title{ Spin-active devices based on graphene / WSe$_2$ heterostructure }
\author{A. Mre\'{n}ca-Kolasi\'{n}ska}
\affiliation{AGH University of Science and Technology, Faculty of Physics and
Applied Computer Science,\\
 al. Mickiewicza 30, 30-059 Krak\'ow, Poland}

\author{B. Rzeszotarski}
\affiliation{AGH University of Science and Technology, Faculty of Physics and
Applied Computer Science,\\
 al. Mickiewicza 30, 30-059 Krak\'ow, Poland}
 
\author{B. Szafran}
\affiliation{AGH University of Science and Technology, Faculty of Physics and
Applied Computer Science,\\
 al. Mickiewicza 30, 30-059 Krak\'ow, Poland}
\begin{abstract}

We consider   graphene on monolayer WSe$_2$ and the spin-orbit coupling induced by the transition-metal dichalcogenide substrate for application to spin-active devices. We study quantum dots and graphene quantum rings as tunable spin filters and inverters. We use an atomistic tight-binding model as well as the Dirac equation to determine stationary states confined in quantum dots and rings. Next we solve the spin-transport problem for dots and rings connected to nanoribbon leads. The systems connected to zigzag nanoribbons at low magnetic fields act as spin filters and provide strongly spin polarized current.
\end{abstract}
\maketitle

\section{Introduction}

Graphene \cite{Neto} has inspired many ideas for applications in electronics and valleytronics and was hoped to be useful in spintronics \cite{ERashba2009,Bercioux2010,Tse2011,Inglot2015}.
%Studies of the Rashba spin-orbit interaction for spintronic devices were performed 
 However, the spin-orbit coupling (SOC) in pristine graphene turns out to be too weak 
%with a tiny  gap of the order of 10 $\mu$eV
 \cite{Gmitra2009} for spintronic applications. %\textmu{}
SOC in graphene can be enhanced by doping \cite{Rocha2010,Sheng2010} or adsorption \cite{Leenaerts2009,Balakrishnan2013,Gmitra2013,Gargiulo2014,Hong2011,Hong2012,Santos2014,Irmer2015,Avsar2015} of  light-element atoms or deposition of heavy ones \cite{Weeks2011,Brey2015}. 
%the deposition of adatoms of hydrogen \cite{Leenaerts2009,Balakrishnan2013,Gmitra2013,Gargiulo2014}, fluorine \%cite{Hong2011,Hong2012,Santos2014,Irmer2015,Avsar2015}, or heavy elements \cite{Weeks2011,Brey2015}. 
However,  the adatoms and dopants introduce disorder,
enhance the scattering, and limit the carrier mobility. To circumvent this problem, SOC can be proximity-induced in graphene combined with two-dimensional transition-metal dichalcogenides (TMDCs) as calculated theoretically \cite{Kaloni2014,Gmitra2015} and shown experimentally \cite{Avsar2014,Wang2015,Wang2016,Yang2016,Volkl2017,Yang2017,Ghiasi2017,Benitez2017,Zihlmann2018,Wakamura2018}.
Upon contact the Dirac point of graphene falls within the energy gap of TMDCs, which preserves the  linear band structure of graphene at the Fermi level \cite{Kaloni2014,Gmitra2015}. 
Moreover, in graphene coupled to TMDCs, the electronic bands acquire a spin texture \cite{Cummings2017} %with an out-of-plane component and an in-plane component perpendicular to the electron momentum,
which can lead to modification of the electron spin, and as a consequence, to spin operations.
%An intriguing phenomenon is the theoretically predicted occurence of quantum spin Hall effect demonstrated for the zigzag ribbons of graphene on WSe$_2$ \cite{Gmitra2016}.

\begin{figure}[tb!]
 \includegraphics[width=\columnwidth]{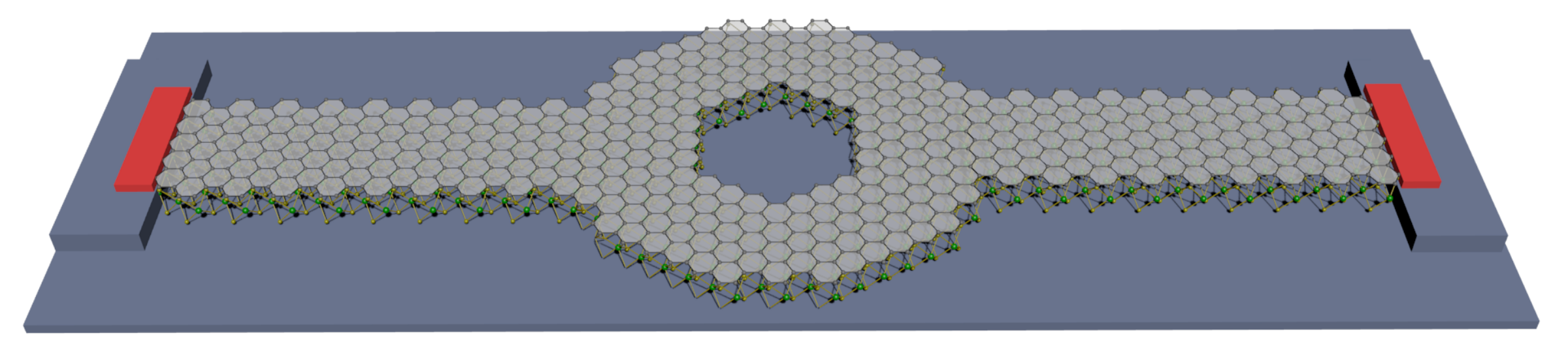}
  \caption{The schematic of the considered system: quantum ring etched out of graphene on WSe$_2$. Two leads are attached to the ring, where electrons enter through the left lead and exit through the right one.
  } \label{systemWSe}
\end{figure}

Previous proposals for spin-active graphene devices, in particular for spin filters, relied on the ferromagnetic substrates \cite{Varykhalov2008,Munarriz2012,JunFeng2012,Rybkina2013} or doping with nonmagnetic atoms \cite{Rocha2010,Sheng2010} like boron, nitride, oxygen, and fluorine. 
However, the proximity of the metal is problematic for the construction of
electronic devices
exploiting the graphene transport properties. In this paper we
consider the graphene/WSe$_2$ heterostructure  with internal
magnetic field due to the spin-orbit coupling instead of the exchange
field, and the substrate has a nonmetallic nature.
We consider using the graphene/WSe$_2$ heterostructure to produce spin-active elements. 
We propose devices built of quantum rings with attached leads
that can be used as a spin filter or spin inverter.
 %The spin filter or spin inverter are of particular interest.
%The in-plane field component can be used for the spin inversion. 
The spin-orbit (SO) interaction for a graphene/TMDC structure implies a perpendicular component of the effective magnetic field of  opposite orientation for both valleys \cite{Cummings2017}. 
The perpendicular component introduces spin splitting of energy bands  provided that the  valley degeneracy is lifted and intervalley scattering is weak. 
The valley degeneracy is lifted by external magnetic field in closed systems, including  quantum rings, dots, and antidots \cite{Thomsen2017,Grujic2011,Recher2007}.
The lifted valley degeneracy with the valley-dependent spin-orbit field leads to the polarization of the spin states. 
Moreover, the in-plane component of the SO effective magnetic field is considered for the spin inversion.
In this work we study a quantum dot and quantum ring
as spin-active elements for the spin currents fed by graphene nanoribbons. We calculate the spectrum of an isolated system and the transport properties of an open system. For the transport calculations we consider a ring with semi-infinite leads attached (Fig.~\ref{systemWSe}).

%
%On the other hand, 
%for spin filtering a splitting of the spin-resolved energy levels is needed -- for this purpose one can use the out-of plane component in Eq.~(\ref{eq:SOfield}), which acts as a Zeeman splitting provided the valley degeneracy is lifted.
%Owing to the last term, due to the intrinsic SOC a coupling between spin and valley occurs. 
%In transport, this property can be used to separate the spins in systems which break the valley degeneracy. 

\section{Theory}
\subsection{Dirac equation}
%\subsection{Model system}
%We consider quantum rings \textcolor{red}{(and quantum dots)} etched out from a graphene contacted to 2-dimensional WSe$_2$. 
We focus on the electronic properties near the Dirac point.
The low-energy Hamiltonian for graphene on TMDCs is \cite{Gmitra2016}
\begin{eqnarray}
  \begin{aligned}
   H =& H_{orb}+H_\Delta+ H_{SO} ,\\
H_{SO}=&H_{I}+H_{R}+H_{PIA} ,
\label{eq:hamLow}
  \end{aligned}
\end{eqnarray}
with
\begin{equation}
  \begin{aligned}
  H_{orb} =& \hbar v_F ( \kappa \sigma_x k_x + \sigma_y k_y ), \\
H_\Delta =& \Delta \sigma_z, 
\label{eq:hamLow2}
  \end{aligned}
\end{equation}
and the SOC terms
\begin{equation}
  \begin{aligned}
 H_{I} =& \frac{1}{2} \left[ \lambda_I^{A}(\sigma_z+\sigma_0) + \lambda_I^{B}(\sigma_z-\sigma_0) \right]\kappa s_z, \\
	H_{R} =& \lambda_R ( \kappa\sigma_x s_y-\sigma_y s_x ), \\
 	H_{PIA}=& \frac{a}{2} \left[ \lambda_{PIA}^A (\sigma_z+\sigma_0) \right.+\\
 	& +\left. \lambda_{PIA}^B(\sigma_z-\sigma_0) \right] ( k_x s_y - k_y s_x ), 
  \end{aligned}
\label{eq:hamLow3}
\end{equation}
where $\kappa=1(-1)$ for the $K$ ($K'$) valley, $\sigma_i$ are the sublattice Pauli matrices, $s_i$ are the spin Pauli matrices, $k_i$ are the components of the wave vector with respect to the $K$ or $K'$ valley, $v_F$ is the Fermi velocity, and $a=0.246$ nm is the graphene lattice constant. The $H_{orb}$ term describes freestanding graphene, and $H_\Delta$ is the staggered potential induced by the TMDC substrate giving rise to an energy gap. $H_I$ describes the intrinsic SOC,  $H_R$ is the Rashba SOC, and $H_{PIA}$ is the pseudoinversion asymmetry \cite{Gmitra2016}.

As shown by Cummings {\it et al.} \cite{Cummings2017}, graphene coupled to a TMDC acquires a spin texture. The effective SO field $\hbar\boldsymbol{\omega}$ can be described by the Hamiltonian written in the basis of the eigenstates of $H_{orb}$,
\begin{equation}
  \begin{aligned}
 H_{SO} &= \frac{1}{2} \hbar \boldsymbol{\omega} \mathbf{\hat s},\\
\hbar \omega_x &= -2(ak \Delta_{PIA} \pm \lambda_R) \sin(\theta), \\
 \hbar \omega_y &= 2(ak \Delta_{PIA} \pm \lambda_R) \cos(\theta), \\
 \hbar \omega_z &= 2 \kappa \lambda_{VZ},
  \end{aligned}
\label{eq:SOfield}
\end{equation}
where $\theta$ is the direction of $k$ relative to $k_x$, $\boldsymbol{\omega}$ is the spin precession frequency, $\Delta_{PIA} = \tfrac{1}{2}( \lambda_{PIA}^{A}-\lambda_{PIA}^{B} )$, $\lambda_{VZ} = \tfrac{1}{2}( \lambda_I^{A}-\lambda_I^{B} )$ parametrizes the valley Zeeman SOC, and the $+(-)$ sign corresponds to the conductance (valence) band. The $\Delta_{PIA}$ and $\lambda_R$ terms contribute to the effective field in the plane of graphene and are both perpendicular to the $k$ vector, while the $\lambda_{VZ}$ term gives an out-of-plane component, which is opposite for the $K$ and $K'$ valleys.

\subsection{Confined states}

For the discussion of the low-energy spectrum of graphene systems we solve the Dirac equation only with the dominant (spin diagonal) terms.
In particular, we focus on the intrinsic SOC, which is responsible for the out-of-plane effective field that leads to the spin polarization in the $z$ direction [see Eq.~(\ref{eq:hamLow3})]:
\begin{equation}
  \begin{aligned}
\nonumber &H' = H_{orb}+H_\Delta +H_{I} . 
\label{eq:hamLowKM}
  \end{aligned}
\end{equation}
The $2\times 2$ Hamiltonian for the valley with the index $\kappa$ takes the form
\begin{equation}
  \begin{aligned}
H_\kappa = \left( \begin{array}{c c}
 \lambda_I^A	\kappa s_z +\Delta	&	\hbar v_F( \kappa k_x -i k_y ) \\ % \otimes \mathbb{I}	
 \hbar v_F( \kappa k_x +i k_y )	&	-\lambda_I^B	\kappa s_z	-\Delta	\\
 \end{array}\right),
%  \left( \begin{array}{c}
%  \psi_1(\mathbf{r}) \\
%  \psi_2(\mathbf{r})   
% \end{array}\right) 
 %=  \varepsilon 
 %  \left( \begin{array}{c}
 % \psi_1(\mathbf{r}) \\
 % \psi_2(\mathbf{r})   
 %\end{array}\right), 
 \label{eq:h2x2}
  \end{aligned}
\end{equation}
where $\mathbf{k}=-i\mathbf{\nabla} + \tfrac{e}{\hbar} \mathbf{A}$, with $\mathbf{A}$ being the vector potential. We use $\mathbf{A}=(-\tfrac{By}{2},\tfrac{Bx}{2},0)$.
 This Hamiltonian acts on a two-component wave function \mbox{$\psi(\mathbf{r})=  \left( \begin{array}{c}
 \psi_1(\mathbf{r}) \\
 \psi_2(\mathbf{r})   
 \end{array}\right)$}, where $\psi_1(\mathbf{r})$ and $\psi_2(\mathbf{r})$ describe the A and B sublattices, respectively. For a circularly symmetric system, these are also eigenstates of the total angular momentum operator $J_z = L_z +\hbar \sigma_z/2$; thus one can  write the wave function as
%\begin{eqnarray}
\begin{equation}
  \begin{aligned}
  \psi(r,\phi) = e^{im\phi}
  \left( \begin{array}{c}
  \chi_1(r) \\
  e^{i\kappa\phi} \chi_2(r)   
 \end{array}\right), 
 \label{eq:wf}
  \end{aligned}
\end{equation}
%\end{eqnarray}
where $m=0,\pm 1,\pm 2,...$ is the total angular momentum quantum number and $\chi_1$ and $\chi_2$ are radial wave functions corresponding to sublattices A and B, respectively. These satisfy the eigenequation 
\begin{widetext}
%\begin{eqnarray}
\begin{equation}
  \begin{aligned}
 E \chi_1 =&  ( \lambda_I^A \kappa \hat s_z + \Delta ) \chi_1 
-i \hbar v_F \left(\kappa \partial_r + \frac{(m+\kappa)}{\rho}+  \frac{Be}{2\hbar} r \right)\chi_2, \\
E \chi_2 =& -i\hbar v_F \left(\kappa\partial_r - \frac{ m}{r}-\frac{Be}{2\hbar}  r \right)\chi_1%(r)
-( \lambda_I^B \kappa \hat s_z + \Delta ) \chi_2 , %(r) 
\label{eq:eqDirak}
  \end{aligned}
\end{equation}
%\end{eqnarray}
\end{widetext}
which we solve with the finite-difference method. %approximation with the backward approximation to the first derivative, as described in \textcolor{red}{cytowanie pracy z Computational? albo arxiv - gdyby byla}.

\subsection{Tight-binding approximation}

For the atomistic modeling, we use the tight-binding Hamiltonian \cite{Gmitra2016}:
\begin{eqnarray}
\nonumber  & H= \sum\limits_{\langle i,j\rangle,s } t c_{i s}^\dagger c_{j s}+\sum\limits_{i,s} \Delta \xi_{c_i} c_{i s}^\dagger c_{is} \\
   &+ \frac{2i}{3} \sum\limits_{\langle i,j\rangle } \sum\limits_{s,s'} c_{i s}^\dagger c_{j s'} \left[\lambda_R ( \mathbf{\hat s} \times\mathbf{d}_{ij} )_z \right]_{s,s'} \\
\nonumber   &+ \frac{i}{3} \sum\limits_{\langle\langle i,j\rangle\rangle } \sum\limits_{s,s'} c_{is}^\dagger c_{j s'} \left[ \frac{\lambda_I^{c_i}}{\sqrt{3}} \nu_{ij} \hat s_z
+ 2\lambda_{PIA}^{c_i} ( \mathbf{\hat s} \times\mathbf{D}_{ij} )_z \right]_{s,s'},
\label{eq:dh}
\end{eqnarray}
where $c_{i s}^\dagger = (a_i^\dagger,b_i^{\dagger} )$ and $c_{is} = (a_i,b_i )$ are the creation and annihilation operators for an electron with spin $s$ in sublattice-A or -B site $i$. The summation $\langle i,j\rangle$ runs over the first-nearest neighbors, and $\langle\langle i,j\rangle\rangle$ runs over the second-nearest neighbors. In the first sum $t$ is the first-nearest-neighbor hopping parameter, and the second sum describes the staggered on-site potential with effective energy difference $\Delta$ with $\xi_{a_i}=1$ ($\xi_{b_i}=-1$) on the A (B) sublattice. The following terms describe the spin-orbit coupling: the Rashba SOC parametrized by $\lambda_R$, the intrinsic SOC term with the lattice-resolved parameter $\lambda_I^{c_i}=\lambda_I^{A(B)}$, and the pseudospin-inversion asymmetry (PIA) with $\lambda_{PIA}^{c_i}=\lambda_{PIA}^{A(B)}$ for $c_i$ in sublattice A(B). $\mathbf{d}_{ij}$ are the unit vectors from site $j$ to $i$ for the nearest neighbors, and $\mathbf{D}_{ij}$ are those for the next-nearest neighbors.
$\mathbf{\hat s}$ is the vector of Pauli matrices acting on the spin state, and
$\nu_{ij}=+ 1 (-1)$ for the clockwise (anticlockwise) path between sites $j$ and $i$.

 %$\nu_{ij}=( \mathbf{d}_{ki}\times \mathbf{d}_{jk} )_z$ \textcolor{red}{with $\mathbf{d}_{ki}$ and $\mathbf{d}_{jk}$ being the unit vectors along the bonds between atoms $k$--$i$ and $j$--$k$, respectively, via their common neighbor $k$.} 
We use the tight-binding parametrization of Ref.~\onlinecite{Gmitra2016} for graphene coupled to WSe$_2$. 
 We introduce the magnetic field by the Peierls phase: a general hopping parameter described by $H=\sum_{i,j,s,s' } h_{i s j s'}  c_{is}^\dagger c_{j s'} $ is modified by $h_{i s js'}\rightarrow h_{is js'}e^{\phi_{ij}}$, where \mbox{$\phi_{ij}= \frac{2\pi i}{\phi_0} \int_{\mathbf{r}_i}^{\mathbf{r}_j} \mathbf{A}\cdot d\mathbf{l}   $}.

\subsubsection{Transport calculation}

\begin{figure}[tb!]
 \includegraphics[width=0.55\columnwidth]{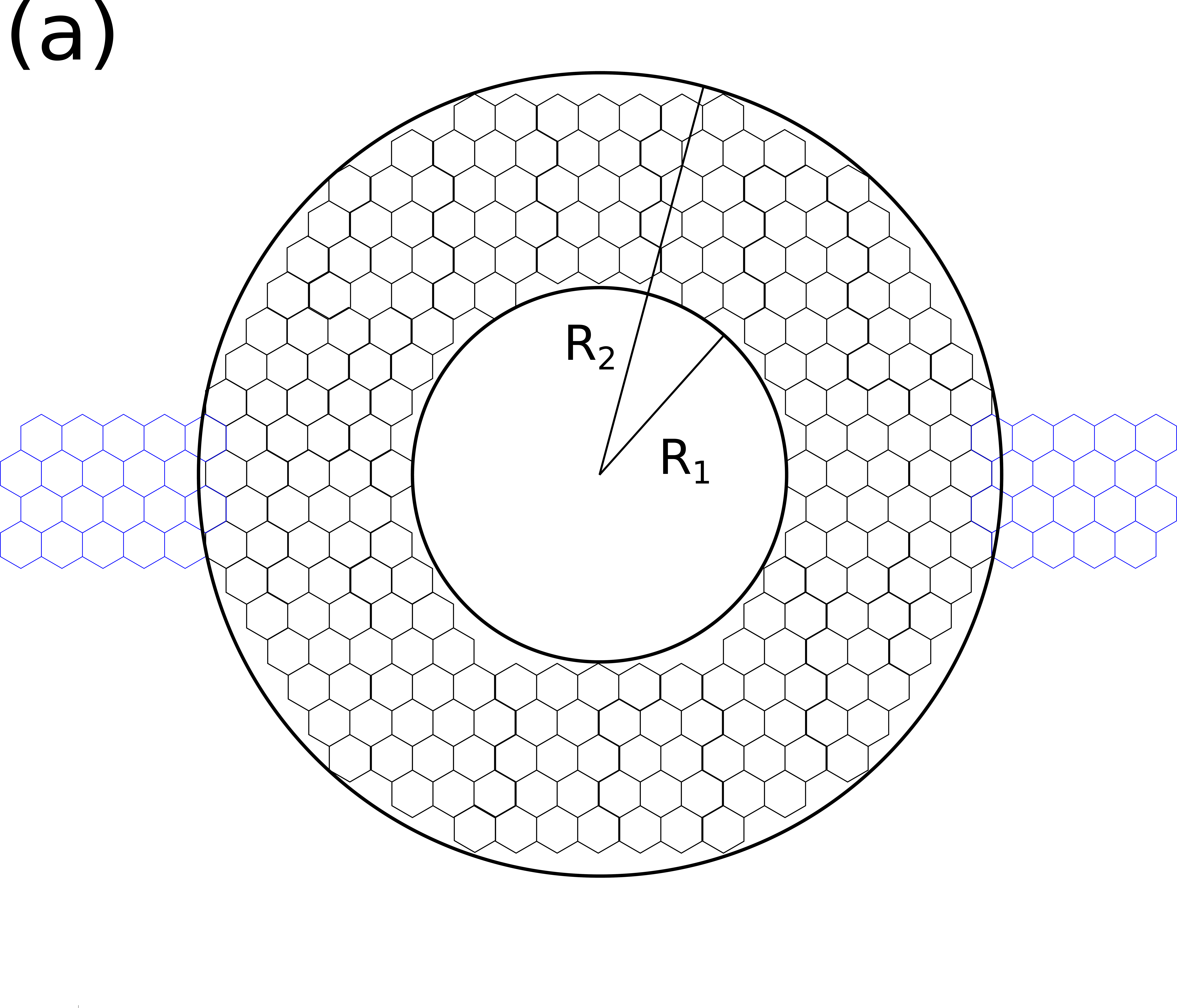}
 \includegraphics[width=0.42\columnwidth]{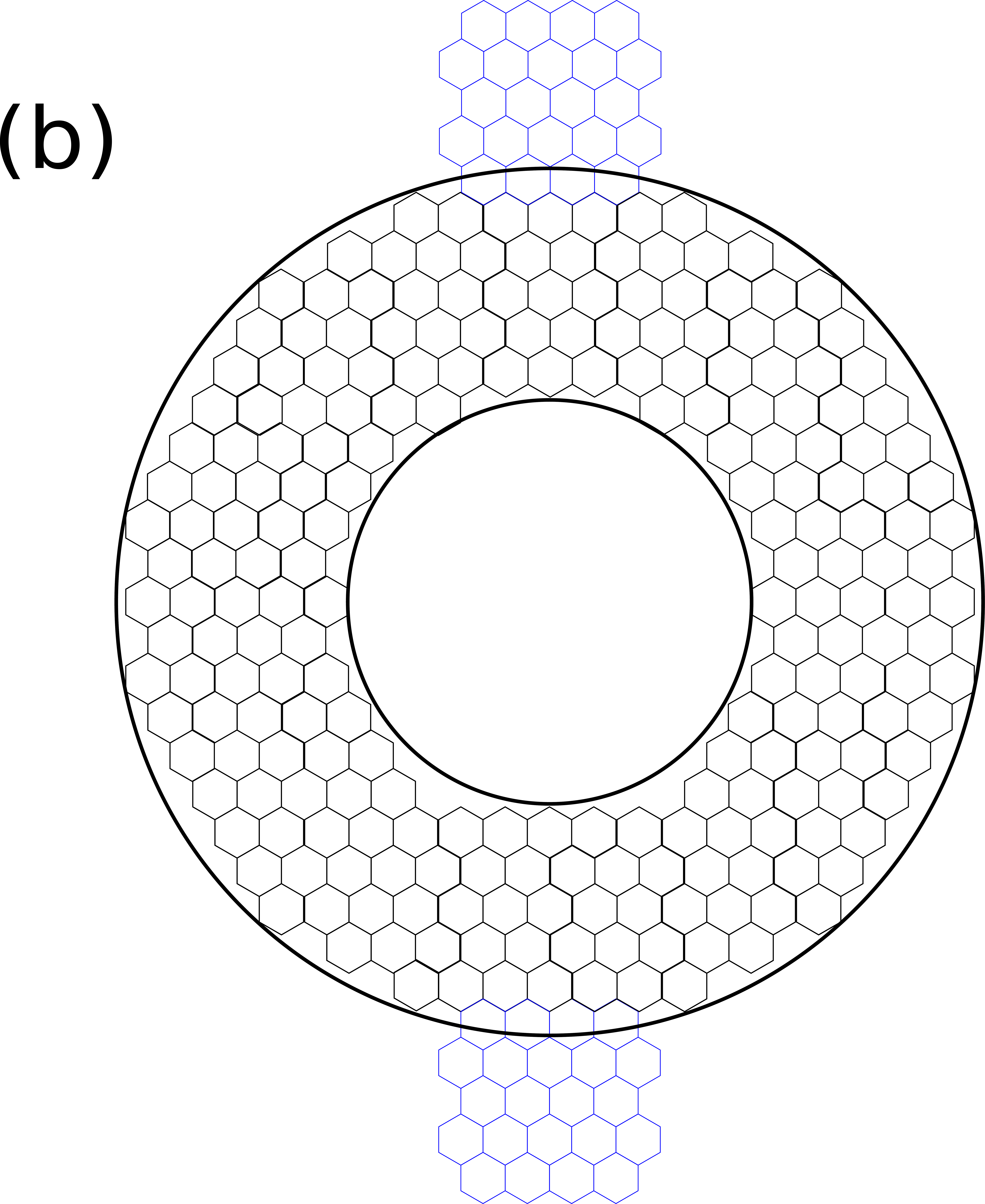}
  \caption{The schematic of the quantum ring with two leads attached with a (a) a zigzag edge and (b) an armchair edge. The zigzag leads are oriented along the $x$ axis, while the armchair leads are oriented along the $y$ axis. The ring inner (outer) radius is $R_1$ ($R_2$).
  } \label{modelTransp}
\end{figure}

We perform the transport calculations in the tight-binding formalism. The considered system is shown in Fig.~\ref{modelTransp}.
It consists of a quantum ring of inner radius $R_1=7.3$ nm and  outer radius $R_2=25$ nm, centered at $(x_0,y_0)=(61.5,25)$ nm.
The quantum ring has two leads attached for the incoming and outgoing electrons. The leads are in the form of narrow ribbons with a zigzag or armchair edge with a width of 17.7 nm, which corresponds to 84 (71) atoms across the zigzag (armchair) ribbon. The edge of the armchair ribbon induces strong intervalley scattering which is absent for the zigzag ribbon.
We maintain the same quantum ring orientation but attach the leads at different angles.
The zigzag leads are oriented along the $x$ axis, while the armchair leads are oriented along the $y$ axis (see Fig.~\ref{modelTransp}). 
We use the gauge appropriate for each terminal using the approach from Ref.~\onlinecite{Baranger1989}. We take $\mathbf{A_0}=(By,0,0)$ and apply the transformation $\mathbf{A}=\mathbf{A_0}+\mathbf{\nabla}[f(x,y)m(x,y)]$, with $f(x,y)=-xyB$, and $m(x,y)$ being a smooth steplike function that is 0 for -5 nm$<y<$55 nm  and 1 elsewhere.

For the evaluation of the transmission probability, we use the wave-function-matching (WFM) technique \cite{Kolacha}. 
The spin direction of the $m$th mode with wave function $\psi^m$ is determined by the quantum expectation values of the Pauli matrices $\langle  \mathbf{\hat s} \rangle = \langle \psi^m | \mathbf{\hat s} | \psi^m \rangle$. We label the positive (negative) spin $\langle \mathbf{\hat s} \rangle$ by $s= \uparrow$ ($\downarrow$).
The transmission probability from the input lead to mode $m$ with spin direction $s$ in the output lead is
\begin{equation}
  \begin{aligned}
T^m_s = \sum_{ n,s' } |t^{mn}_{s,s'}|^2,
\label{eq:transprob}
  \end{aligned}
\end{equation}
where $t^{mn}_{s,s'}$ is the probability amplitude for the transmission from mode $n$ with spin $s'$ in the input lead to mode $m$ with spin direction $s$ in the output lead. The summed transmission to spin $s$ is
\begin{equation}
T_s = \sum_{ m } T^m_s.
\label{eq:transprob_spin}
\end{equation}
We evaluate the summed conductance as $G={G_0}\sum_{s} T_s$, with $G_0={e^2}/{h}$.
We consider the spin-conserving $G_{ss}$ and spin-flipping $G_{ss'}$ components, given respectively by $G_{ss} = G_0 \sum_{ n,m } |t^{mn}_{s,s}|^2$ and $G_{ss'} = G_0\sum_{ n,m } |t^{mn}_{s,s'}|^2$.
For discussion of the spin filtering we use the spin polarization defined as 
\begin{equation}
P = \frac{ G_{\uparrow\uparrow}+G_{\downarrow\uparrow} - (G_{\downarrow \downarrow}+G_{\uparrow\downarrow}) }{G}.
\label{eq:polarization}
\end{equation}
For the system filtering out the spin-up (spin-down) electrons this gives $P=-1$ ($P=1$).
The transport properties below are discussed within the energy range in which
only subbands with opposite spin appear at the Fermi level. The orientation of the
spin depends on the type of the ribbon feeding the current to the lead and on the
external magnetic field. The $\uparrow$ and $\downarrow$ in formula (\ref{eq:polarization}) and
below in the discussion stand for the orthogonal spin eigenstates which depend on the case.

%\begin{equation}
%T_{\sigma\sigma'} = \sum_{ n,m } |t^{mn}_{\sigma,\sigma'}|^2.
%\label{eq:transprob_conserving}
%\end{equation}

%For the calculation of the current between the $m$ and $n$ atoms, we use the formula derived from the Schr\"odinger equation \cite{Wakabayashi}:
%\begin{equation}
% {\bf J}_{mn} =  \frac{i}{\hbar} \left[ t_{mn} \Psi^*_m \Psi_n - t_{nm} \Psi^*_n \Psi_m \right],
%\label{current}
%\end{equation}
%where $ \Psi_n $ is the wave function at the $n$th atom.

%The probability current flux can be evaluated as:
%\begin{equation}
% \phi = \sum\limits_m \sum\limits_{n_m}  J_{mn_m} ,
%\label{currentFlux}
%\end{equation}
%where the first sum runs over the atoms at the desired cross-section of the ribbon, and the second sum over their neighbors $n_m$ localized to the right.

\section{Results}

\begin{figure}[tb!]
 \includegraphics[width=0.95\columnwidth]{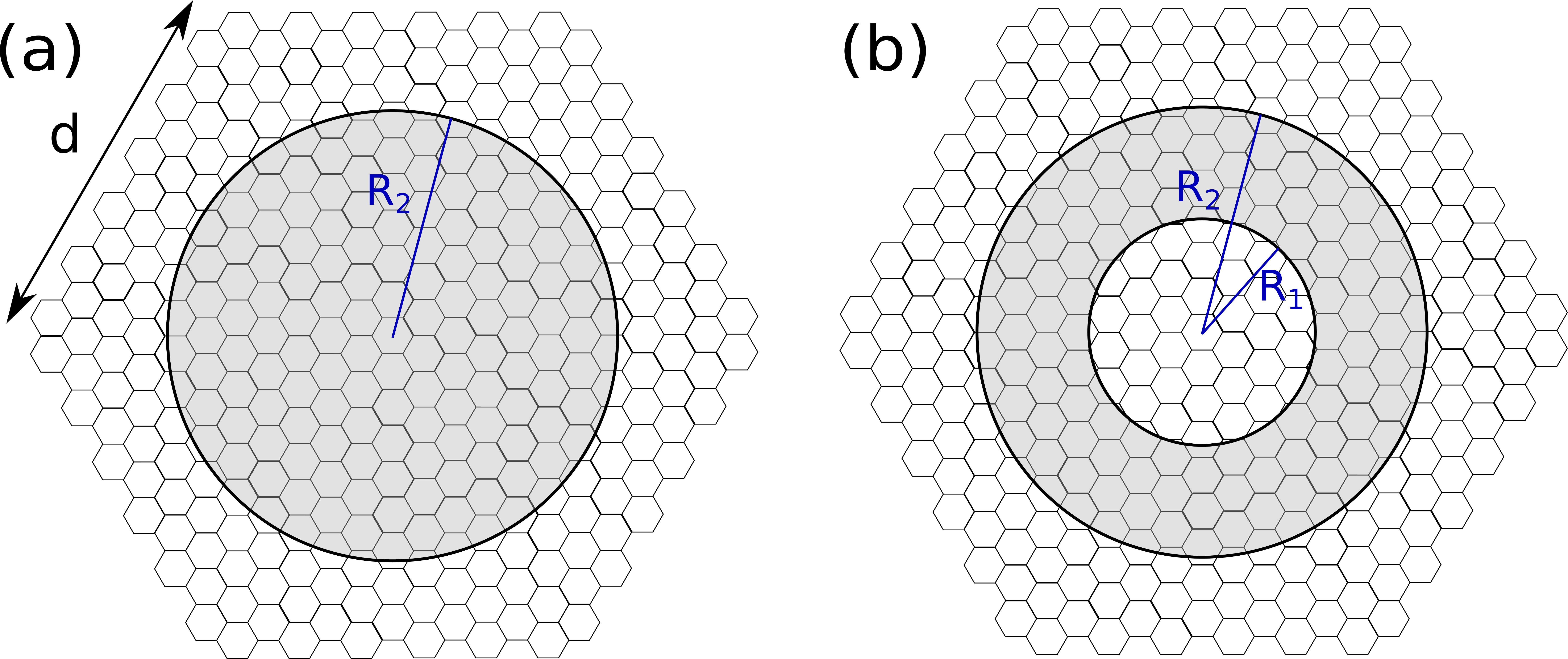}
  \caption{The schematic of the (a) dot and (b) ring induced by a staggered potential in the hexagonal graphene flake with side length $d$. In the shaded area the staggered potential is zero, and outside of it the on-site potential is given by Eq.~(\ref{eq:deltaDot}) for the dot and (\ref{eq:deltaRing}) for the ring. The dot radius is $R_2$, and the ring inner (outer) radius is $R_1$ ($R_2$).
  } \label{schemeIMBC}
\end{figure}

\subsection{Effective mass-induced closed quantum ring}

\begin{figure*}[tb!]
  \includegraphics[width=0.46\textwidth]{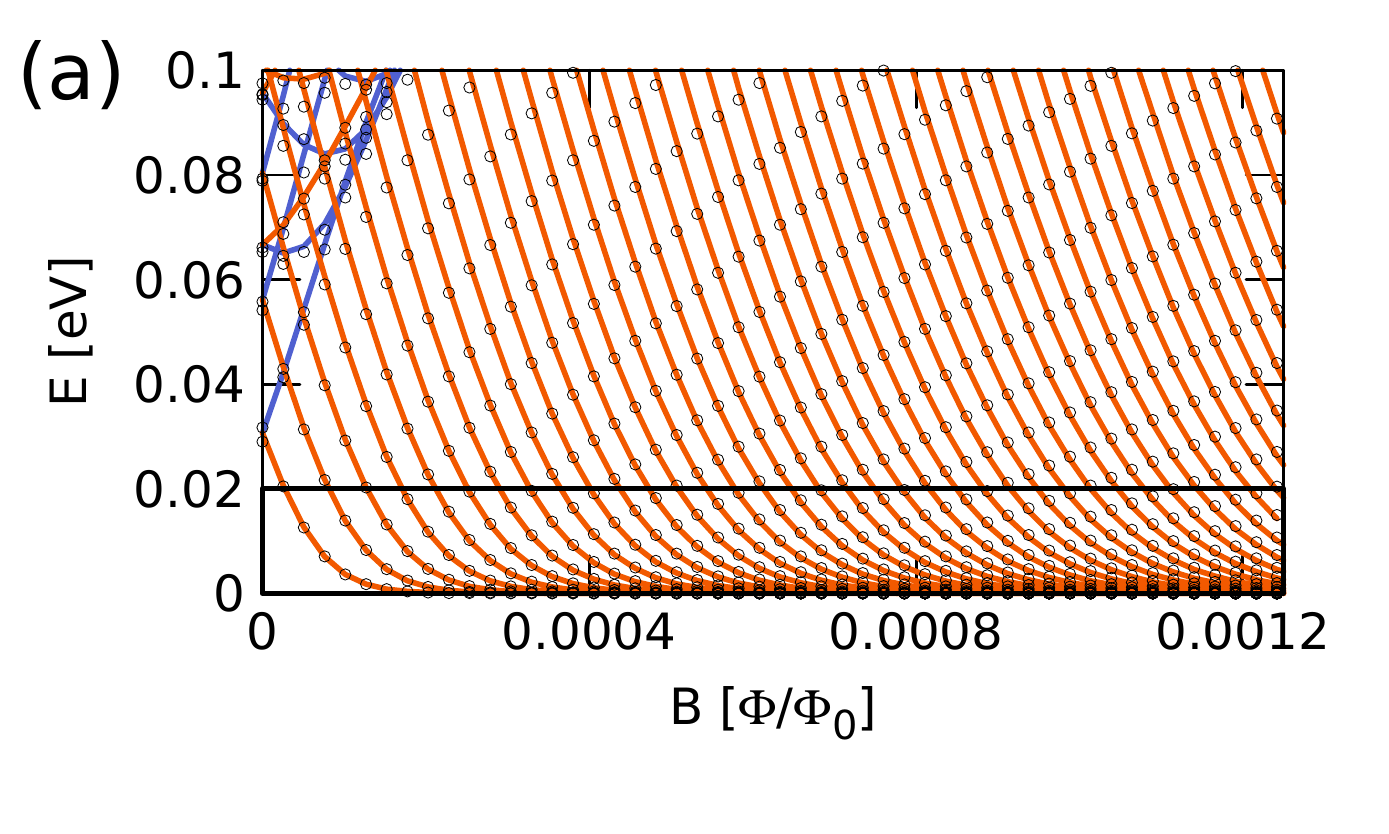}
  \includegraphics[width=0.46\textwidth]{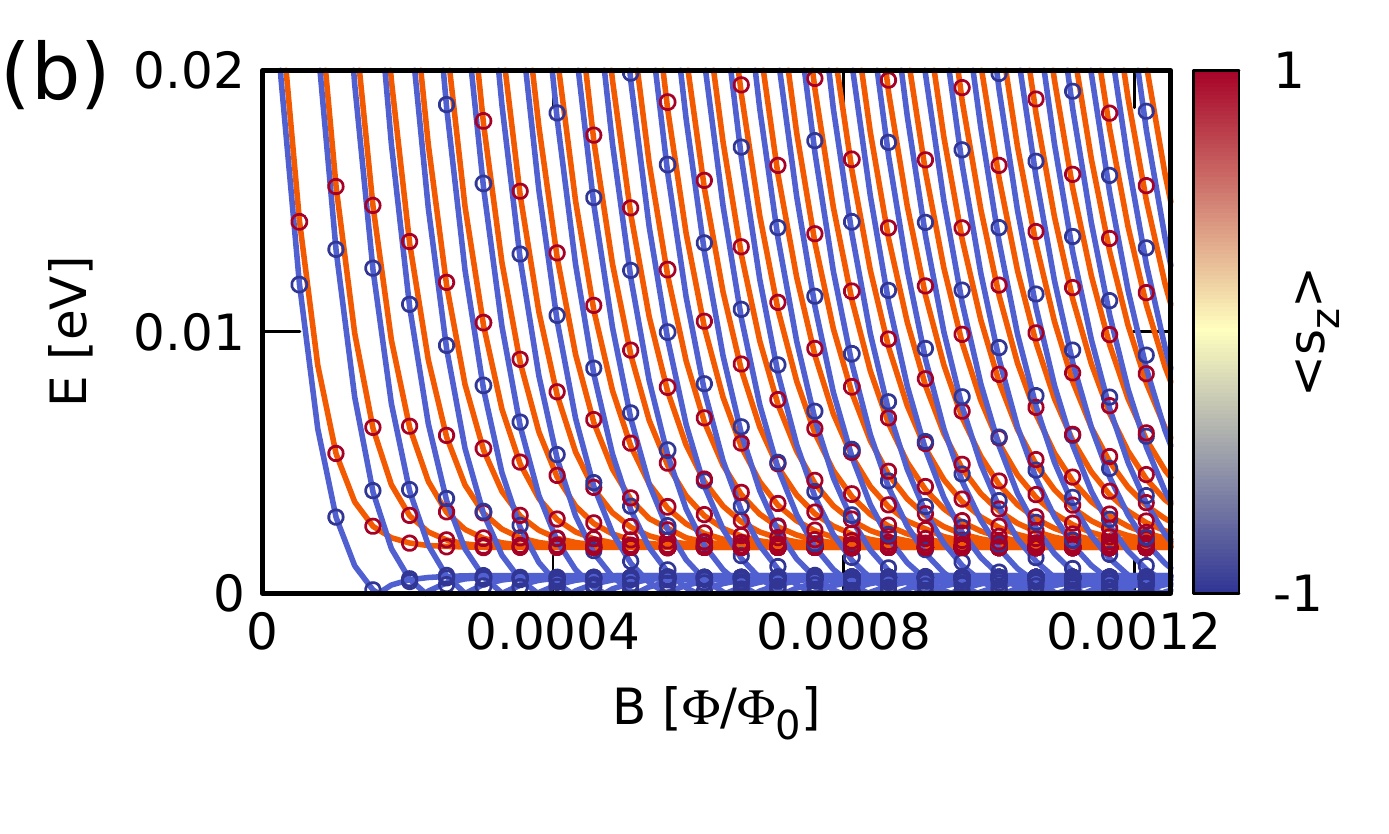}
  \includegraphics[width=0.46\textwidth]{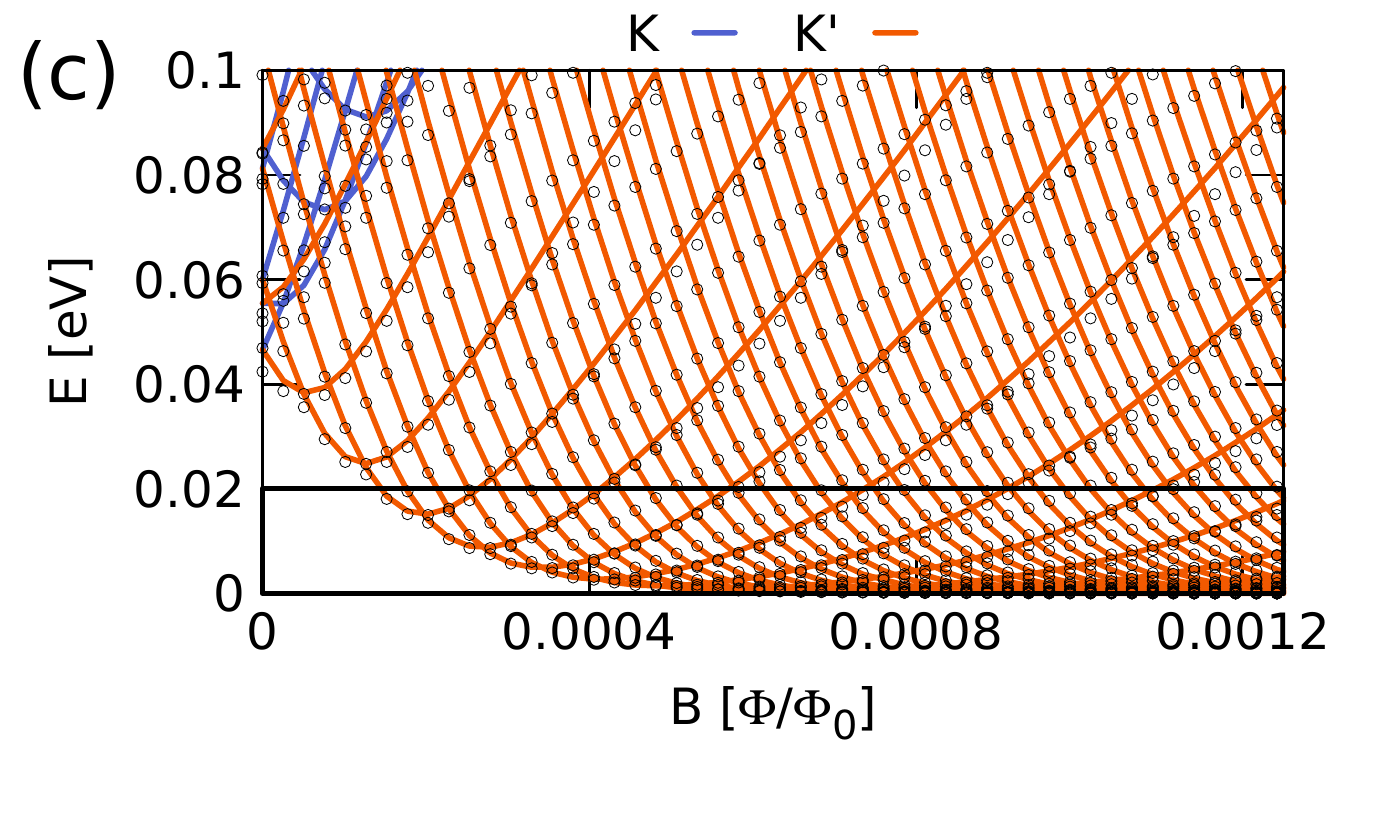}
  \includegraphics[width=0.46\textwidth]{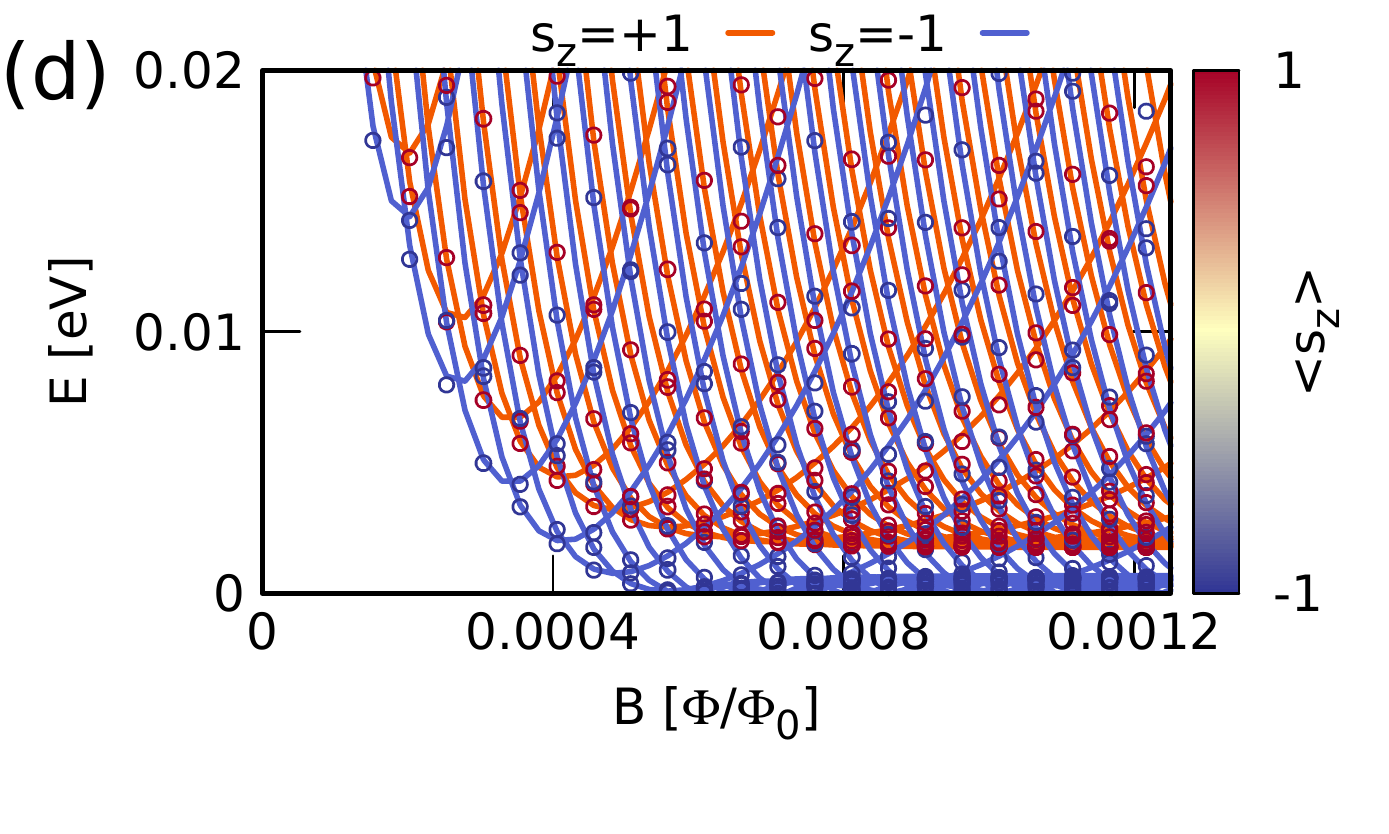}
  \caption{The low-energy spectrum of the electrostatic potential-induced quantum ring and dot (a) and (c) of suspended graphene and (b) and (d) with the proximity-induced SOC as a function of external magnetic field, with the tight-binding spectrum drawn by dots with the color scale showing the $z$ component of the spin. The lines show the continuum approximation spectrum. The black rectangles in (a) and (c) show where the zooms in (b) and (d) are taken, respectively. The colorscale is the same for (a,c) and (b,d).
  } \label{eigenDiracTB}
\end{figure*}

As a proof of concept for the spin filtering by quantum dots and rings in external magnetic field, we focus on the low-energy properties of graphene systems with proximity-induced spin-orbit coupling.
For this purpose, we calculate the spectrum of Hamiltonian (\ref{eq:hamLowKM}) in the continuum approximation for a quantum dot and ring defined by infinite-mass boundary conditions \cite{Berry53}. The considered dot has a radius of $R_2=25$ nm, and the ring has an outer radius $R_2=25$ nm and inner radius $R_1=7.3$ nm.

For comparison, we calculate the tight-binding spectrum of an analogous system presented schematically in Fig.~\ref{schemeIMBC}, where the dot or ring is defined by a staggered potential $\delta( r )$ that introduces a mass term:
\begin{equation}
  H= H_{orb}+H_\Delta+H_I
+ \sum\limits_{i,s} \delta( r) \xi_{c_i} c_{is}^\dagger c_{is}.
\label{eq:dhKM}
\end{equation}
For the confined states we consider only the intrinsic SOC, and the radial potential $\delta( r )$ for the dot is
\begin{equation}
  \begin{aligned}
\delta(r) = \left\{ \begin{array}{c c}
V_g, & r>R_2,\\
0,	& r<R_2,
\end{array}\right.
\label{eq:deltaDot}
  \end{aligned}
\end{equation}
and for the ring it is
\begin{equation}
  \begin{aligned}
\delta(r) = \left\{ \begin{array}{c c}
V_g, & r>R_2,\\
0,	& R_1<r<R_2,\\
V_g, & r<R_1,
\end{array}\right.
\label{eq:deltaRing}
  \end{aligned}
\end{equation}
with $V_g=5 $ eV. The staggered-potential-defined system, dot or ring is defined in a hexagonal graphene flake with a side length $d=$30.5 nm (see Fig.~\ref{schemeIMBC}).
 The results for the continuum and tight-binding approximations are presented in Fig.~\ref{eigenDiracTB}.
In Figs.~\ref{eigenDiracTB}(a) and \ref{eigenDiracTB}(c) for pristine graphene without SOC in the continuum approximation, one can determine the valley to which the levels belong. At $B=0$ the energy levels are valley degenerate [see orange (blue) lines for $K$ $(K')$ valley in Figs.~\ref{eigenDiracTB}(a) and \ref{eigenDiracTB}(c)] and in finite magnetic field the levels split for different valleys.
The energy levels obtained in the continuum approximation and tight-binding approach agree, especially for a low energy range and high magnetic field. 

In Figs.~\ref{eigenDiracTB}(b) and \ref{eigenDiracTB}(d) the spectrum for systems with intrinsic SOC is shown. The levels are no longer spin degenerate in both the continuum and tight-binding approximations. %, and the energy splitting is same in both spectra. 
For levels of the $K'$ valley spin-down states are lower in energy than spin-up states due to the out-of-plane valley-Zeeman SOC field. For the $K$ valley this field has the opposite sign; thus the spin-up states have lower energy. % [comp. eg. Fig.~\ref{eigenDiracTB}(a,c)].
The energy splitting of the levels due to intrinsic SOC for WSe$_2$ is $E_{VZ} = 2\hbar \omega_{z} = 2.38$ meV.

In the systems confined by the position-dependent mass the intervalley scattering
is absent, and the valley degeneracy is lifted by a finite magnetic field. 
The effective SO magnetic field $\hbar \omega_z$ is activated in Eq.~(\ref{eq:SOfield}), and due to the $\kappa$ term, it is opposite for both valleys. 
However, in systems that contain sections of armchair edges, mixing of both valley states occurs, which leads to a reduction of the effective magnetic field and the resulting spin splitting (see the next section).

\begin{figure*}[tb!]
 \includegraphics[width=0.485\textwidth]{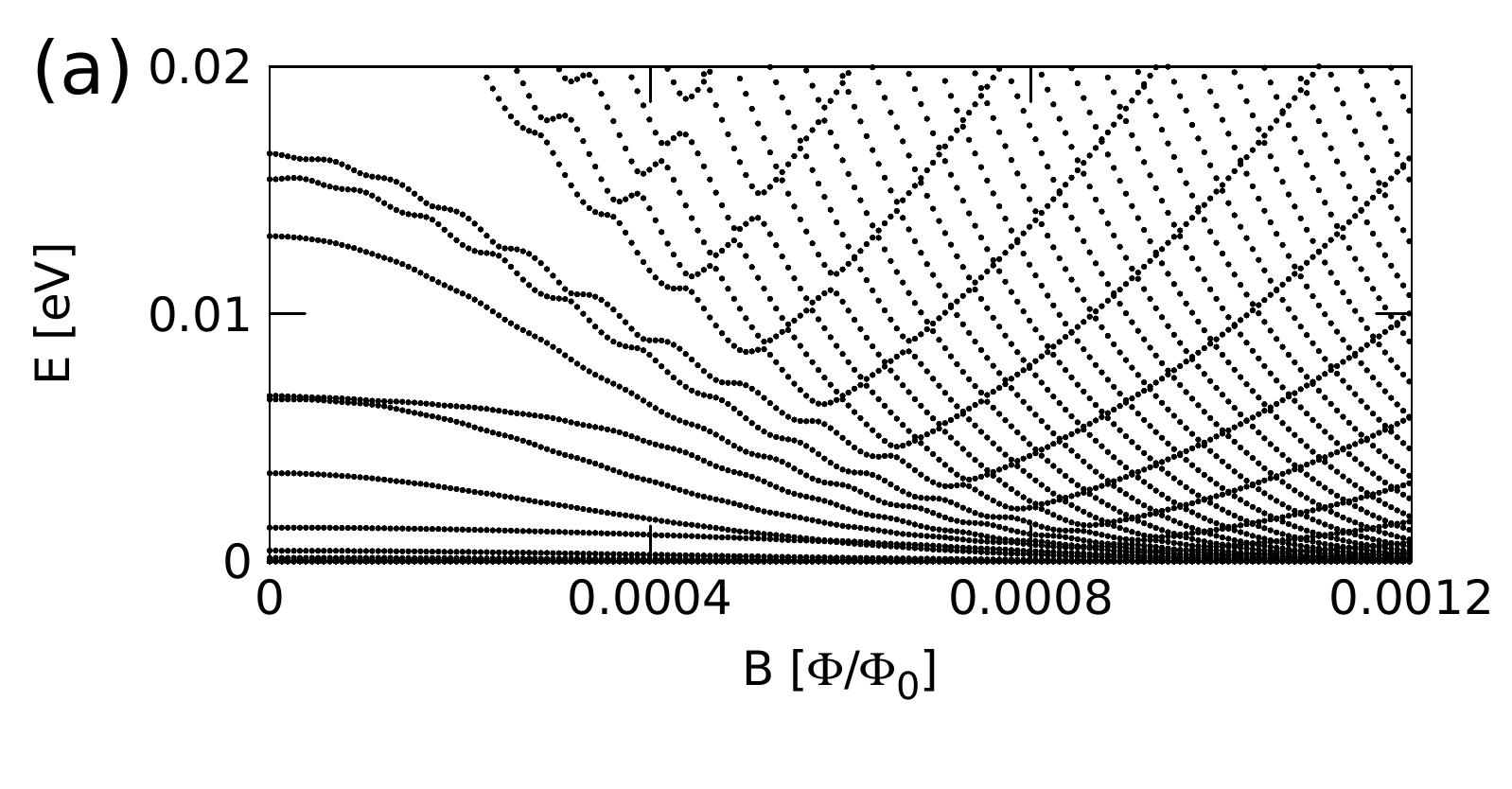}
  \includegraphics[width=0.5\textwidth]{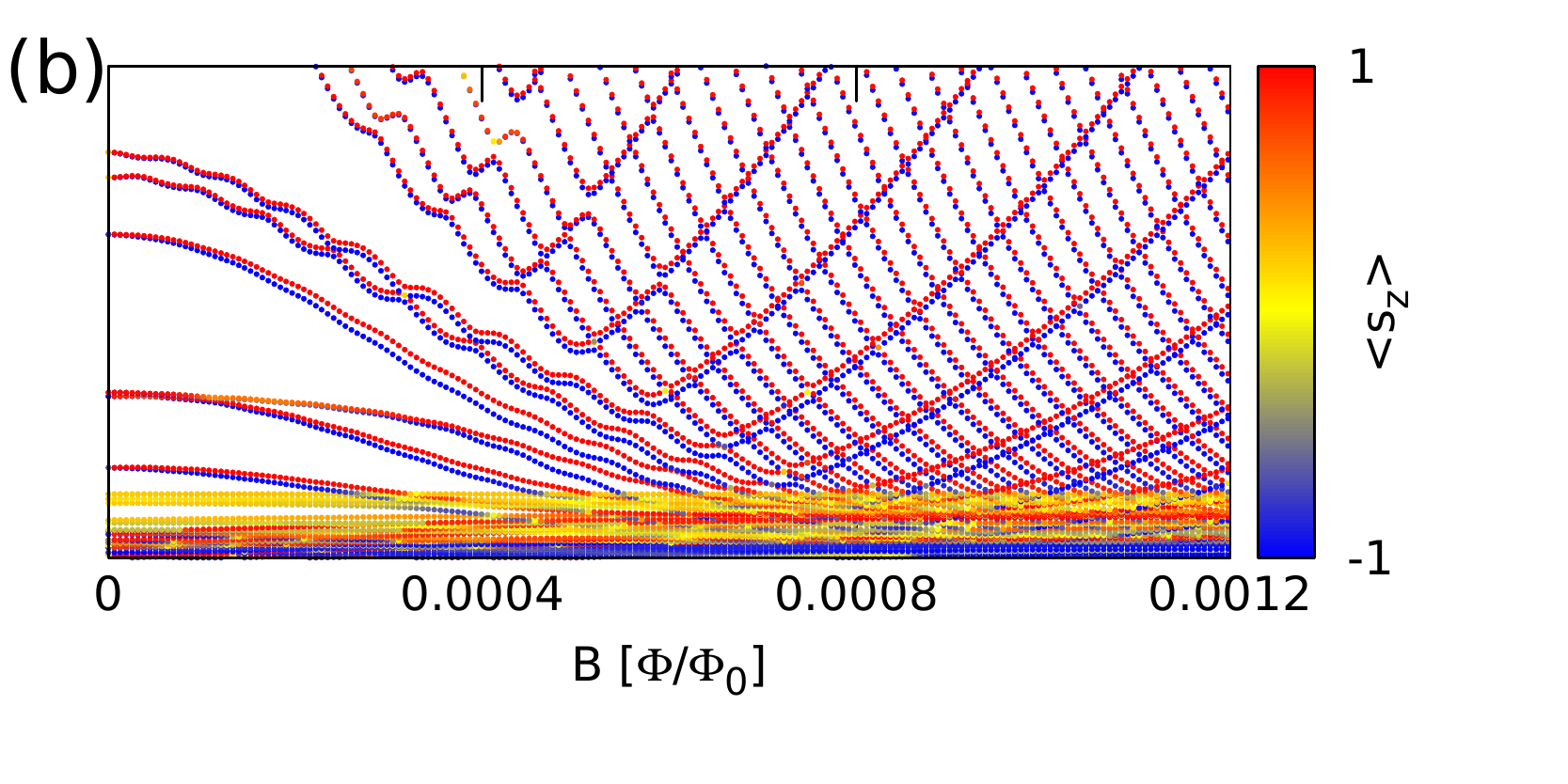}
  \caption{The eigenenergies of the etched quantum ring (a) of pure graphene and (b) with the proximity-induced SOC as a function of external magnetic field. The color scale shows the $z$ component of the spin.
  } \label{eigenring}
\end{figure*}

\begin{figure*}[tb!]
  \includegraphics[width=0.525\textwidth]{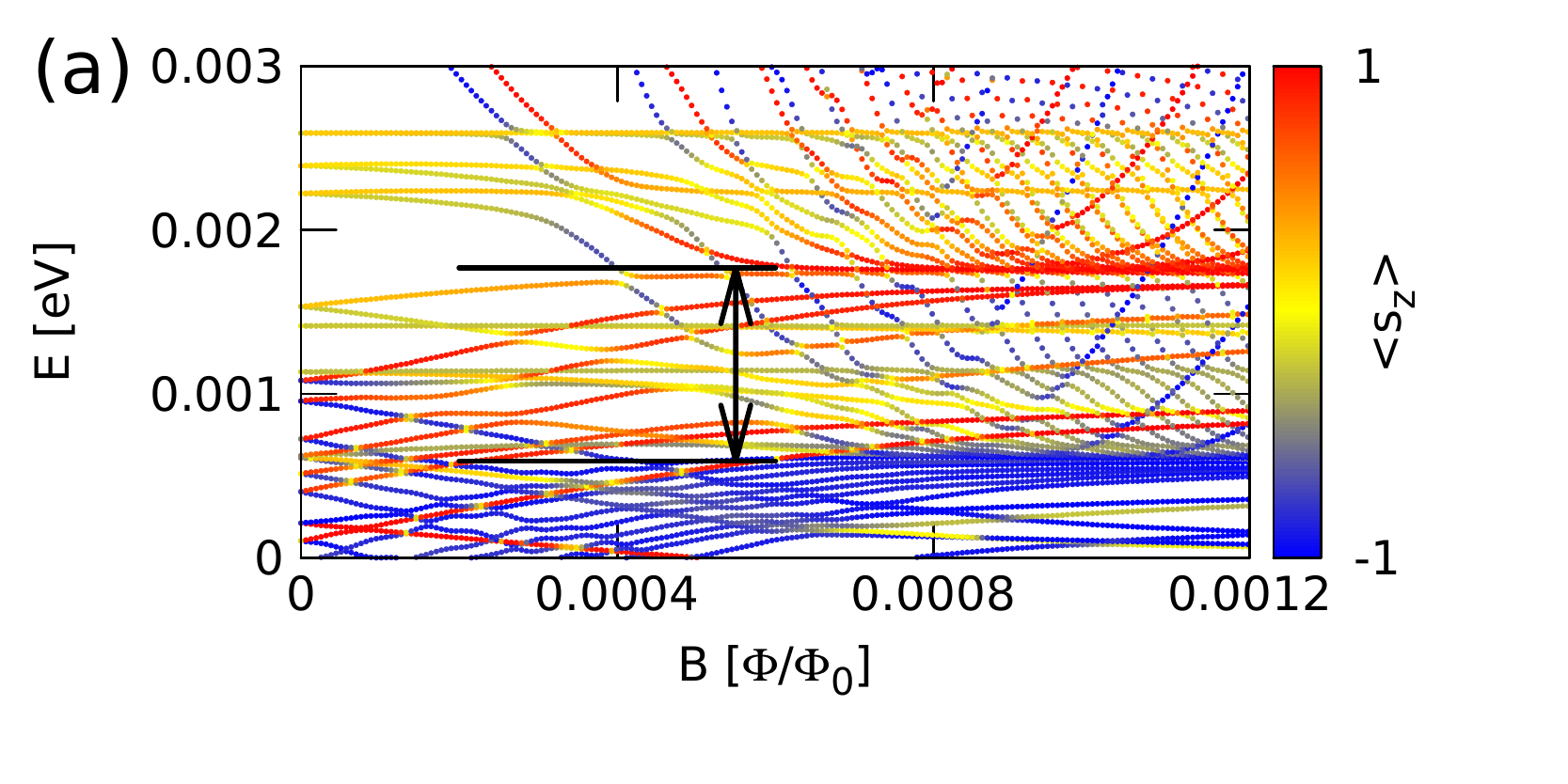}
  \includegraphics[width=0.46\textwidth]{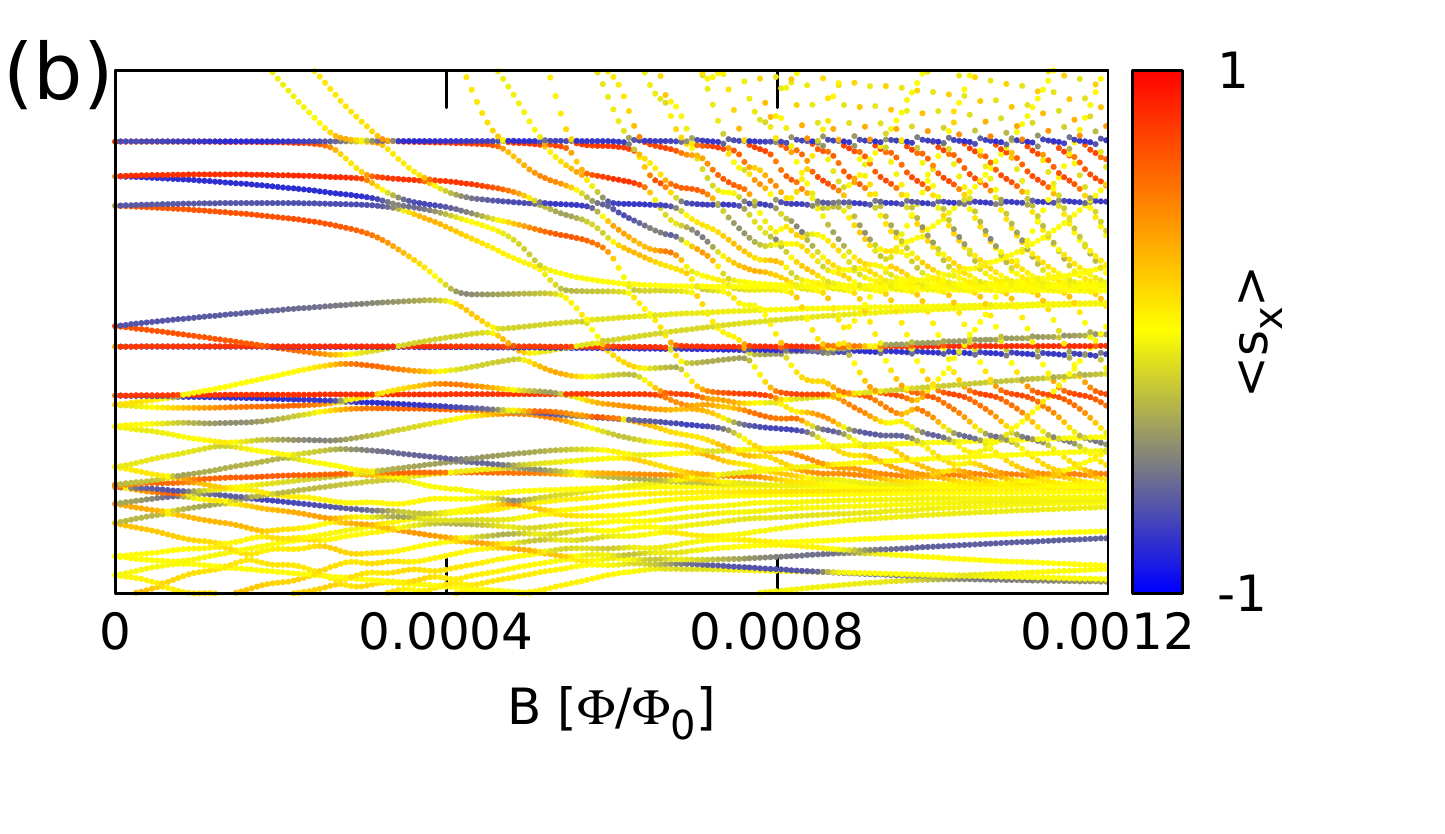}
  \caption{The low-energy spectrum of the etched quantum ring with the proximity-induced SOC as a function of external magnetic field, with the color scale showing (a) $z$ component and (b) $x$ component of the spin.
  } \label{eigenringLow}
\end{figure*}

\subsection{ Closed quantum ring with etched edges}

We now turn our attention to etched systems. In order to reduce the numerical cost of the tight-binding calculations, we focus on quantum rings, which have a smaller number of atomic sites than the quantum dots. However, quantum rings and dots will have similar spin-splitting properties, the dots having the advantage of a smaller disorder caused by etching.

 Here the simple continuum model with finite-mass confinement does not apply since the edge contains short zigzag and armchair sections. Due to the zigzag edge, states close to zero energy can occur, in comparison with the results with infinite-mass boundary conditions, which generate no zero-energy states.

The ring has inner radius $R_1=7.3$ nm and  outer radius $R_2=25$ nm.
In Fig.~\ref{eigenring} we present the spectra of the ring as a function of magnetic flux through one carbon hexagon $\phi/ \phi_0$ (with $\phi_0=h/e$ and $\phi=3\sqrt{3}a_{CC}^2 B$, with $a_{CC}=0.142$ nm) in suspended graphene [Fig.~\ref{eigenring}(a)] and graphene deposited on WSe$_2$ [Fig.~\ref{eigenring}(b)]. Figure \ref{eigenringLow} shows the low-energy zoom of the spectra.

The spectrum is different from the one obtained with the finite-mass confinement. We note that in contrast to the finite-mass quantum dots and rings [Figs.~\ref{eigenDiracTB} (a) and \ref{eigenDiracTB}(c)], for low magnetic field the spectrum of the etched system contains energy levels close to zero (Fig.~\ref{eigenring}). Such states were shown in Ref.~\onlinecite{Grujic2011} to be localized in the zigzag segments of the quantum dot edges. The staggered potential produces armchair and zigzag boundaries, but they do not act as a physical edge and do not support the zero-energy levels. Moreover, it was shown in Ref.~\onlinecite{Grujic2011} that the spectrum in the etched systems strongly depends on the details of the edge structure, whereas in the finite-mass-induced systems it is less sensitive to the imperfections of the circular shape.

For high energies the spectra in the ring of suspended graphene and graphene on WSe$_2$ are very close to each other. 
The spectrum has two series of periodic levels, with energies decreasing (growing) with magnetic field, which correspond to the states associated with a clockwise (anticlockwise) current in the ring and are localized near the outer (inner) edge of the ring \cite{Bahamon2009,Poniedzialek} and having angular momentum parallel (antiparallel) to the external magnetic field \cite{Poniedzialek,MrencaLorentz}. In both series of states the spin is parallel to the $z$ direction.%, unless the two spin-states are nearly degenerate. 
The width of those states is of the order of 10 nm at $\phi=0.0008 \phi_0$ and decreases with growing magnetic field. 

For low energy (below approximately 0.01 eV) the spectra start to differ. Various effects can be seen that result from different SOC terms: 
%Apparently, $\lambda_R$ polarize the spin the $x-y$ plane. Also $\lambda_{PIA}^{A,B}$ contribute to the polarization, .
$\lambda_{I}^{A,B}$ causes the spin polarization out of plane, and splits the levels of opposite spin in energy. % due to the Zeeman-like action. 
On the other hand, due to the terms dependent on $\lambda_R$ and $\lambda_{PIA}^{A,B}$ an in-plane spin component arises. This is especially pronounced at low energy. %[see Fig.~\ref{eigenringLow}(b)].

Figure~\ref{eigenringLow} shows the low-energy zoom of the spectrum in Fig.~\ref{eigenring}. The eigenstates have spin predominantly in the $z$ direction [Fig.~\ref{eigenringLow} (a)], with the exception of the states with energy weakly dependent on the magnetic field close to the Dirac point. 
In the former case the spin is in the $x-y$ plane [Fig.~\ref{eigenringLow}(b)], which suggests that those states %acquire the spin texture due to 
are governed by 
the Rashba-like SOC terms. These states are mostly localized on the segments of the ring that contain the Klein edge \cite{KleinEdge, He}.

The states that carry clockwise or anticlockwise current, also for low energy, have spin almost perfectly polarized in the $z$ direction. The opposite spin levels are split in energy 
due to the $\lambda_{VZ}$ term in Eq.~(\ref{eq:SOfield}). The splitting of the order of 1.14 meV can be seen in Fig.~\ref{eigenringLow}, as highlighted by the black arrow, lower than the maximum of $2.38$ meV because the ring contains short armchair segments which lead to intervalley scattering and, as a consequence, to partial cancellation of the energy splittings of the two valleys.

\subsection{ Magnetotransport of the quantum ring }
\subsubsection{Zeeman splitting neglected}

\begin{figure}[tb!]
%\begin{center}
 \includegraphics[width=0.47\columnwidth]{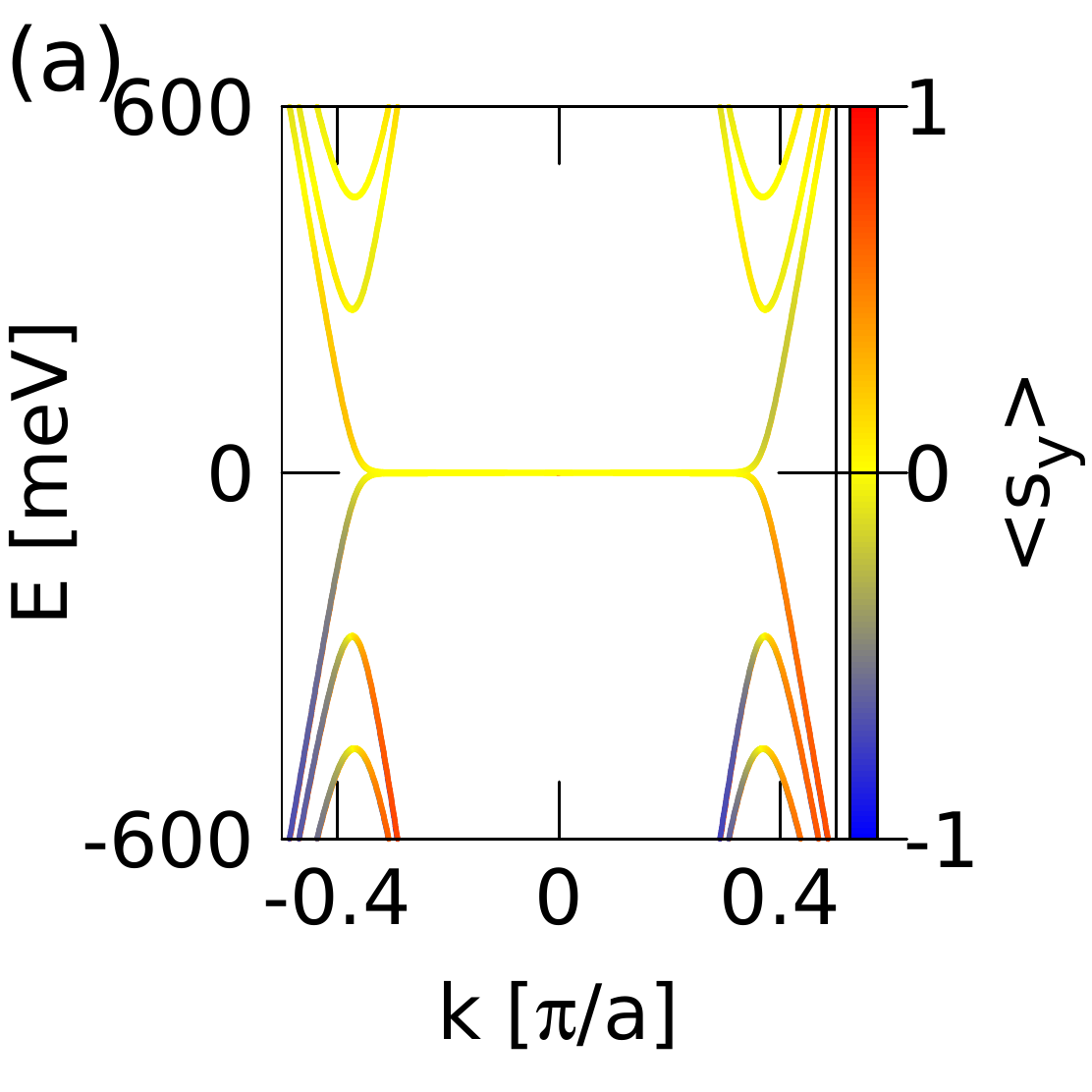}
 \includegraphics[width=0.4\columnwidth]{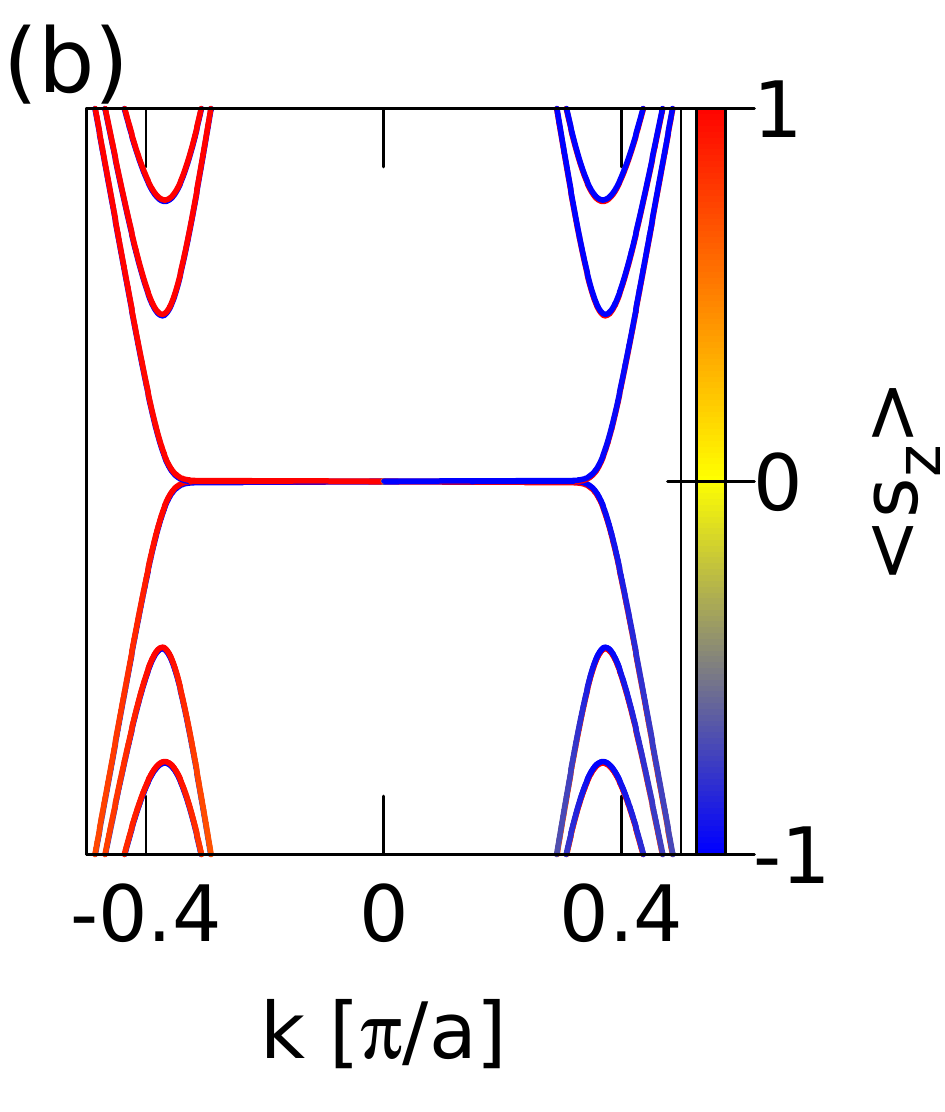}
 \includegraphics[width=0.47\columnwidth]{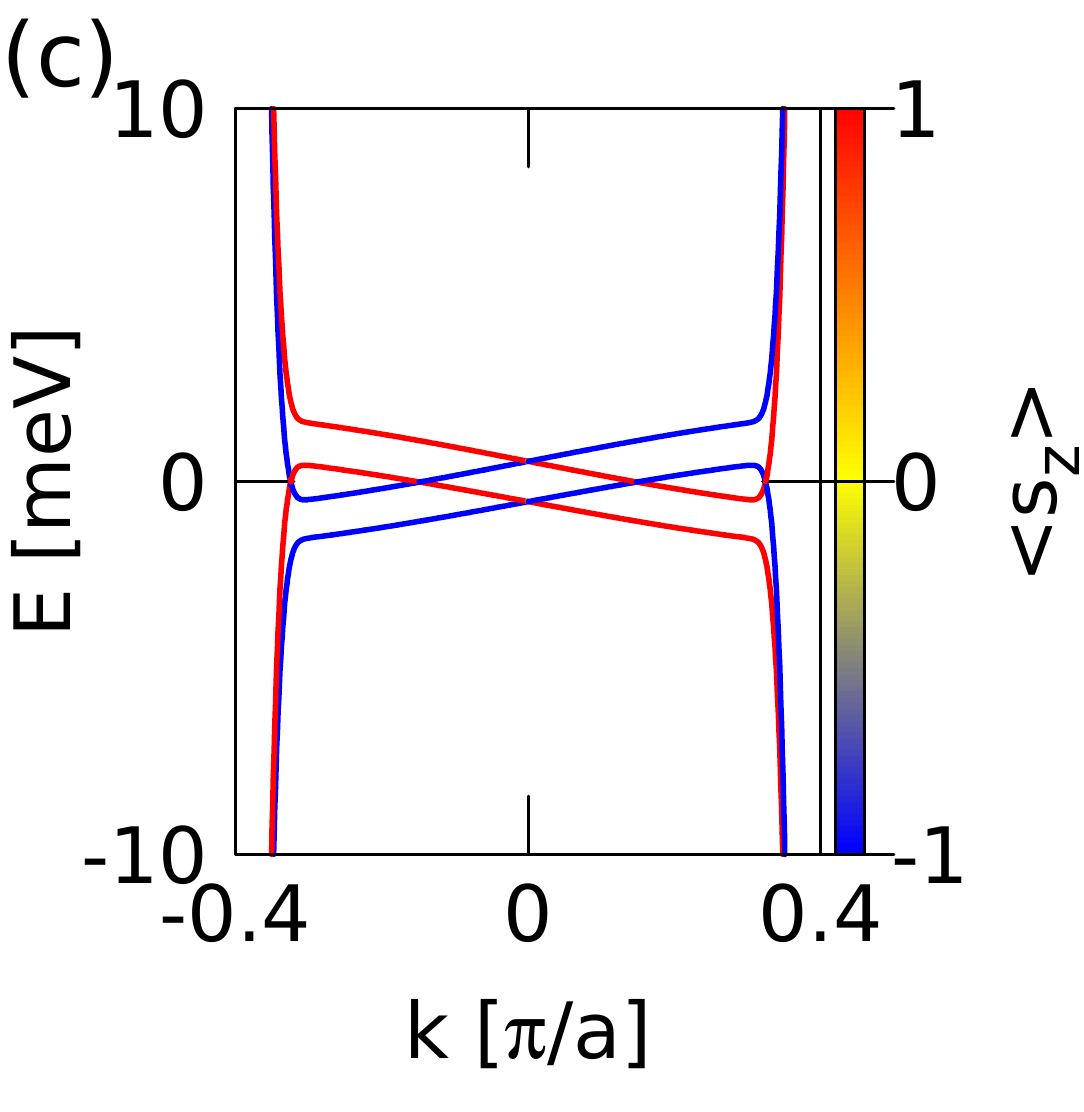}
 \includegraphics[width=0.4\columnwidth]{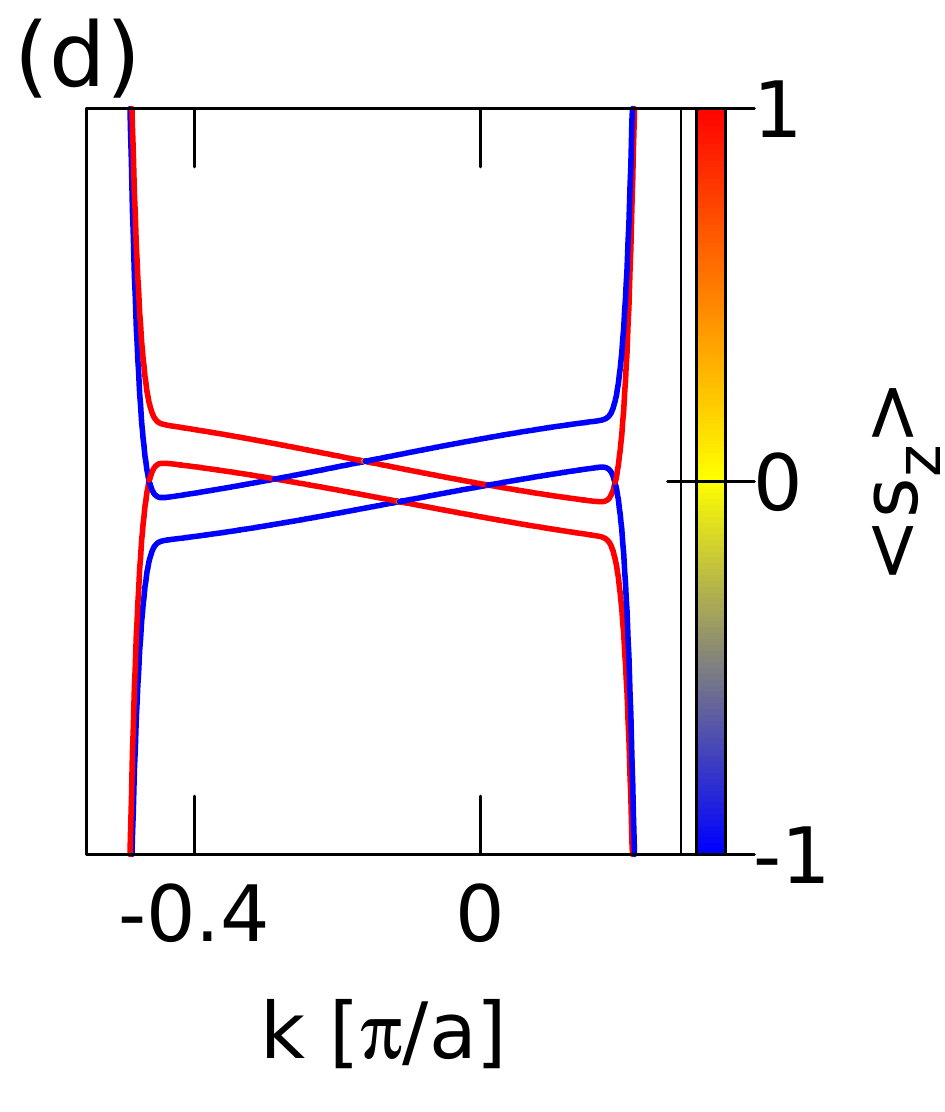}
%\end{center}
  \caption{The band structure of the zigzag nanoribbon with WSe$_2$ for (a)-(c) $\phi=0$ and (d) $\phi=0.0005\phi_0$. The color scale shows (a) the mean spin $y$ or (b)-(d) $z$ component.
  } \label{dispRibbon}
\end{figure}

\begin{figure}[tb!]
 \includegraphics[width=0.5\columnwidth]{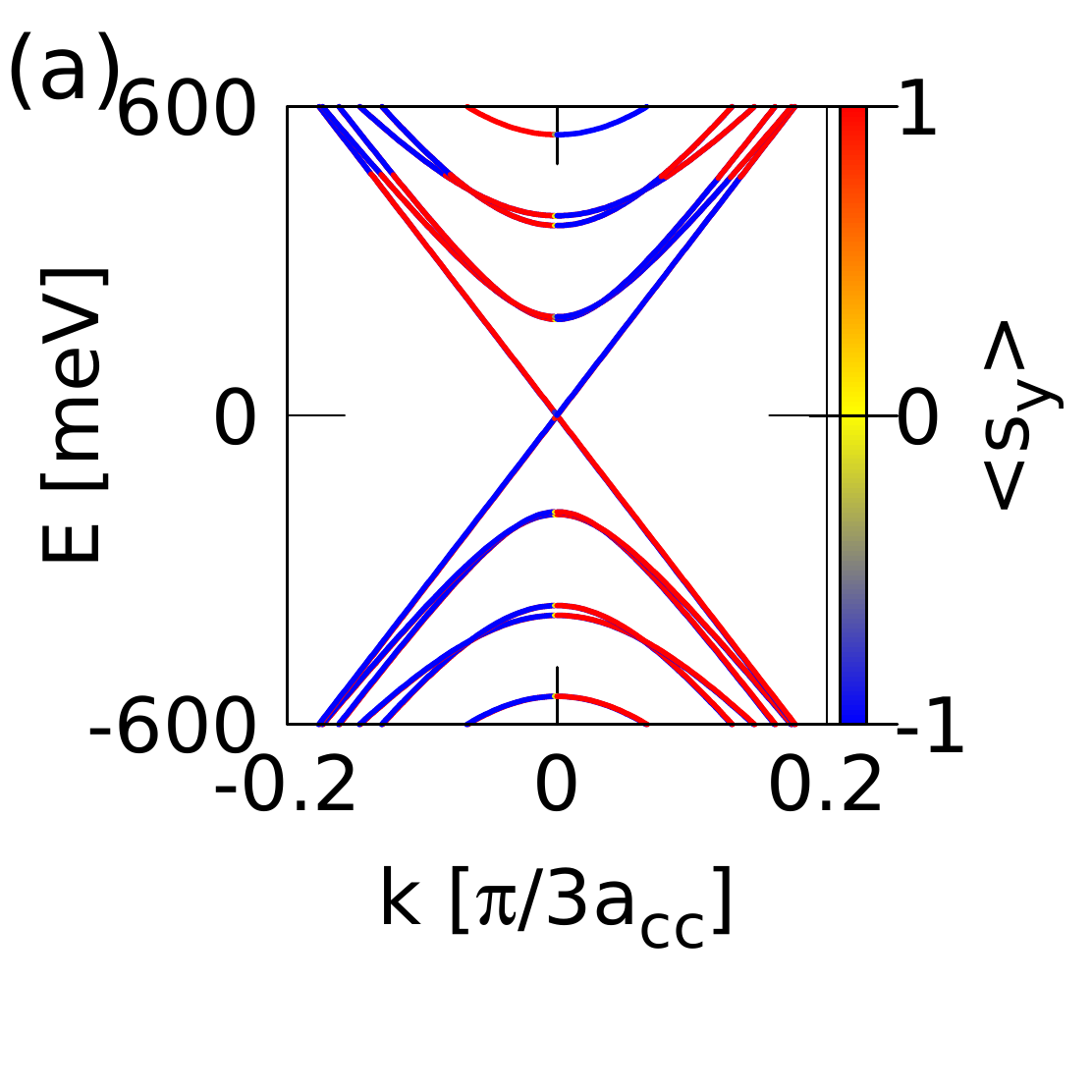}
 \includegraphics[width=0.48\columnwidth]{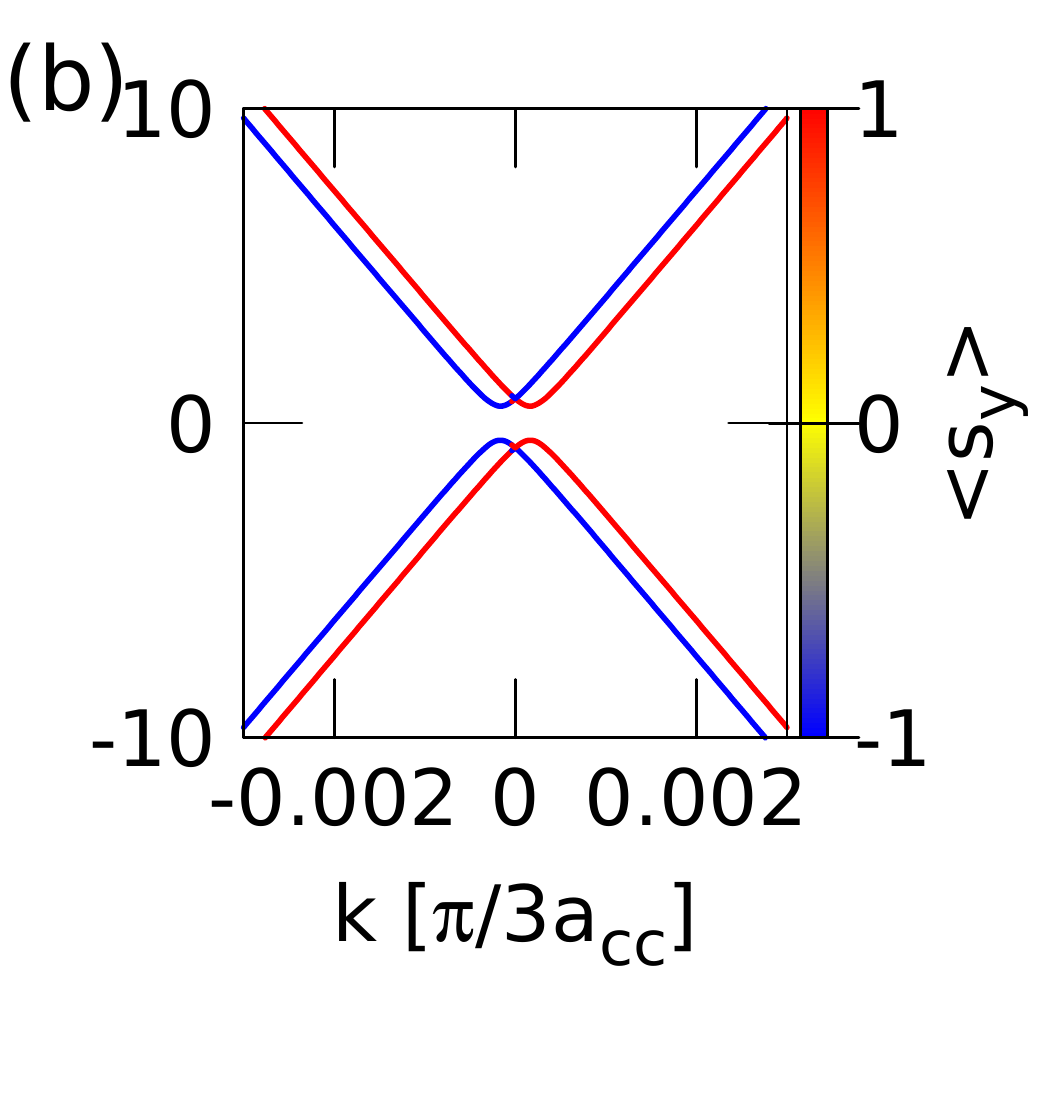} 
 \includegraphics[width=0.5\columnwidth]{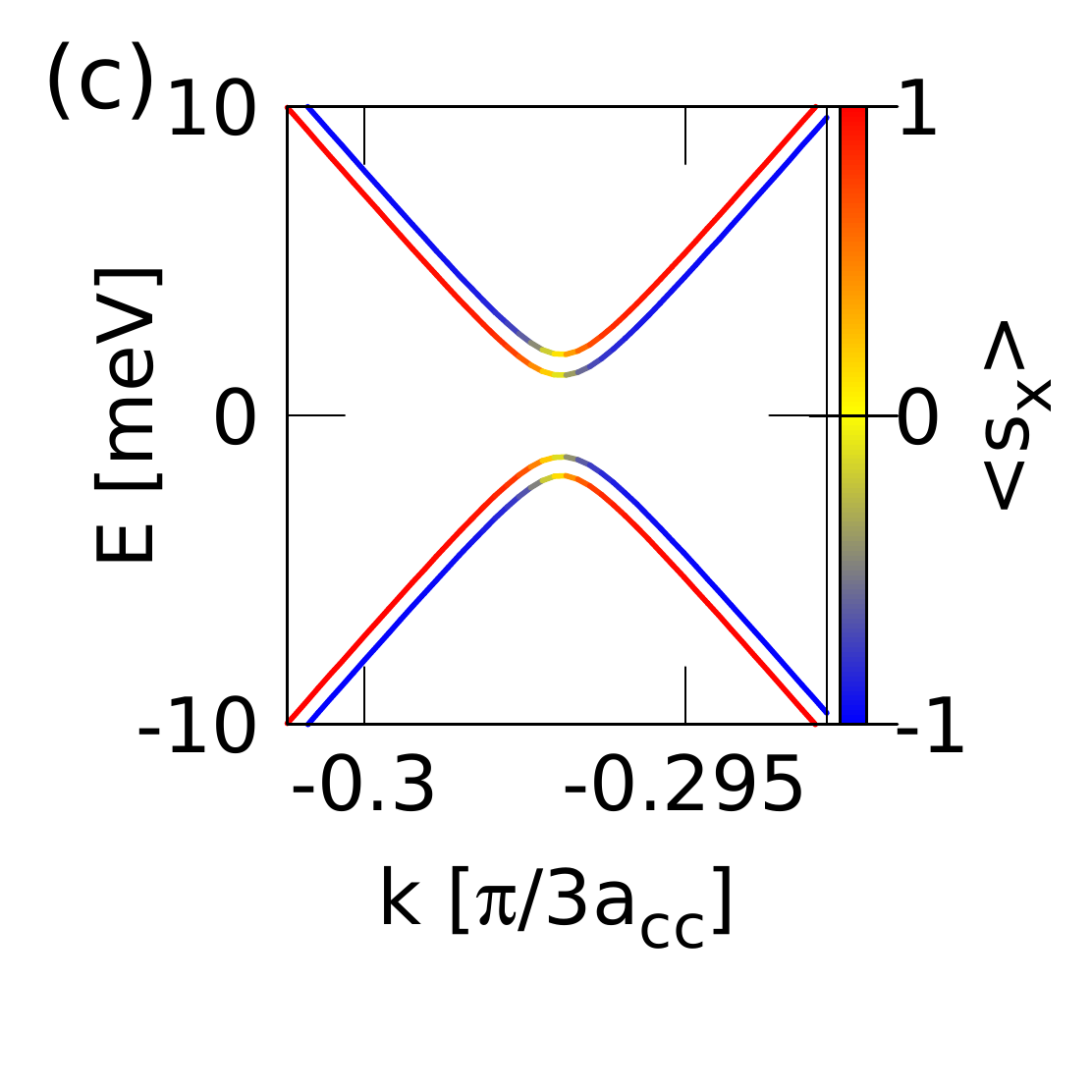}
 \includegraphics[width=0.48\columnwidth]{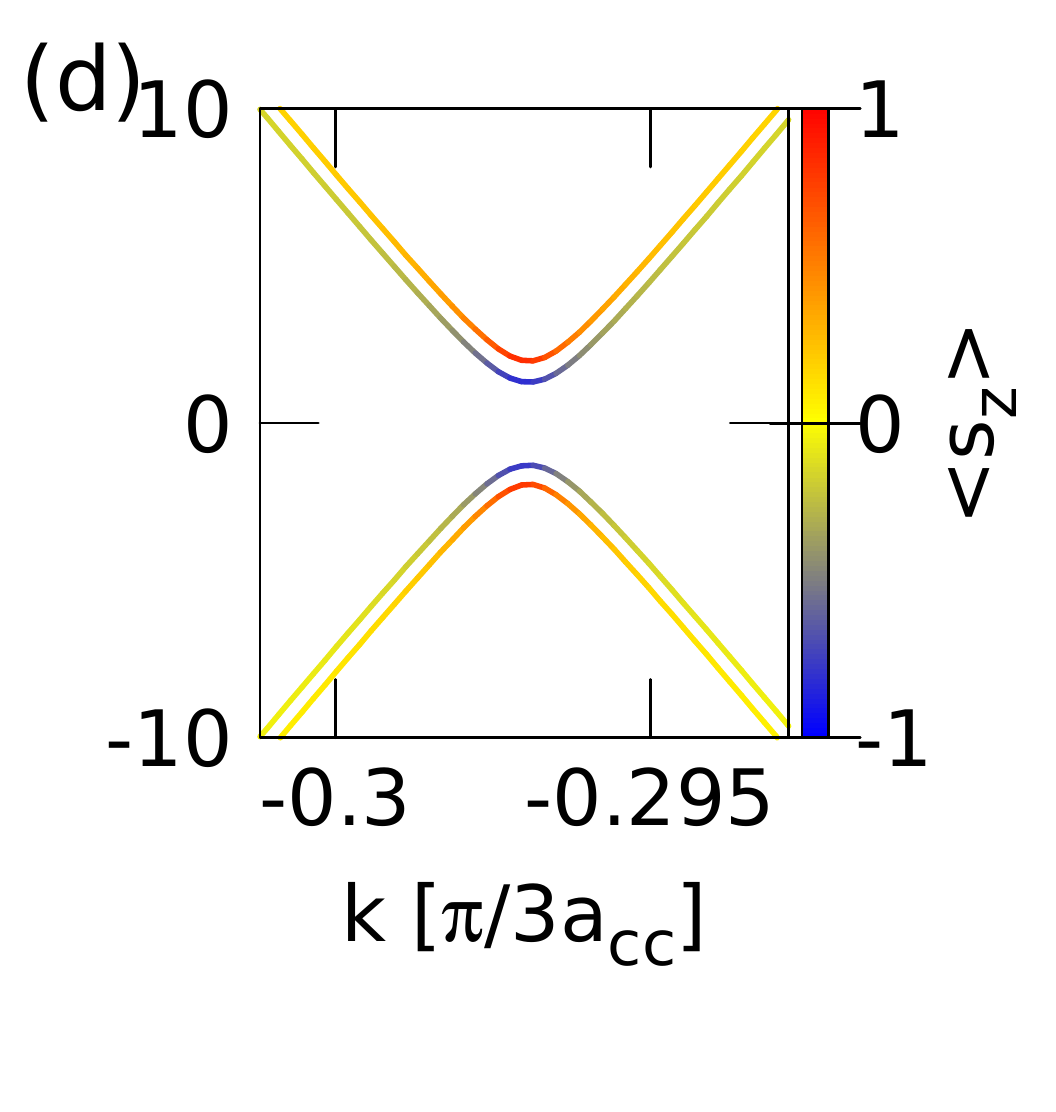}
  \caption{The band structure of the armchair nanoribbon with WSe$_2$ for (a) and (b) $\phi=0$ and (c) and (d) $\phi=0.0005 \phi_0$.
  } \label{dispRibbonArm}
\end{figure}

In this section we present the results of the transport in a quantum ring the same size as in the previous section with the leads attached.
The spin transport depends on the properties of the ring and the leads. From formula (\ref{eq:SOfield}) it is evident that the $z$ component of the SOC field is opposite for the two valleys. It affects the spin direction depending on the edge type of the graphene system. An armchair edge leads to the intervalley scattering, whereas for a zigzag edge the valleys are well defined \cite{pcc,Wurm2012}. Figures \ref{dispRibbon} and \ref{dispRibbonArm} show band structures of zigzag and armchair nanoribbons of graphene on WSe$_2$, respectively. For the zigzag nanoribbon (Fig.~\ref{dispRibbon}) in the dispersion relation the $K$ and $K'$ valleys are around $k_x=\pm2\pi/3a$. The bands have spin aligned almost perfectly in the $z$ direction [Figs.~\ref{dispRibbon}(a) and \ref{dispRibbon}(b)], with the spin direction in the lowest subband being opposite for the two valleys [Fig.~\ref{dispRibbon}(c)]. This is still the case in finite external magnetic field [Fig.~\ref{dispRibbon}(d)].

The armchair edge, on the other hand, mixes valleys, and in zero external field the contributions of the spin field for the $K$ and $K'$ valleys cancel out. Thus the spin is polarized in the nanoribbon plane (Fig.~\ref{dispRibbonArm}), perpendicular to the direction of motion. In Figs.~\ref{dispRibbonArm}(a) and \ref{dispRibbonArm}(b) the bands are clearly polarized in the $x$ direction (for the nanoribbon aligned along the $y$ axis). Only in finite external magnetic field does the spin get tilted out of plane [Figs.~\ref{dispRibbonArm}(c) and \ref{dispRibbonArm}(d)].

\begin{figure*}[tb!]
 \includegraphics[width= 0.49\textwidth]{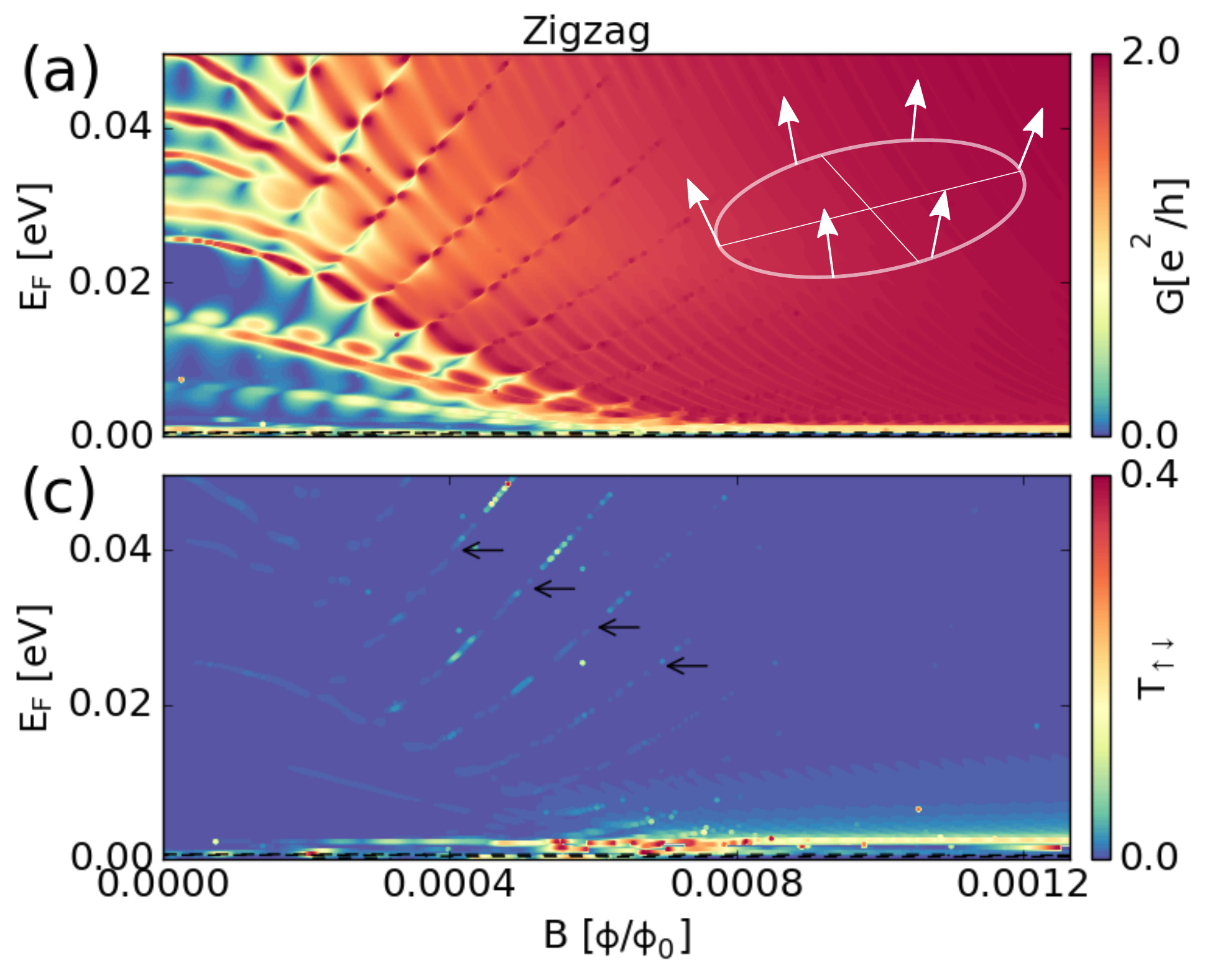}
 \includegraphics[width= 0.5\textwidth]{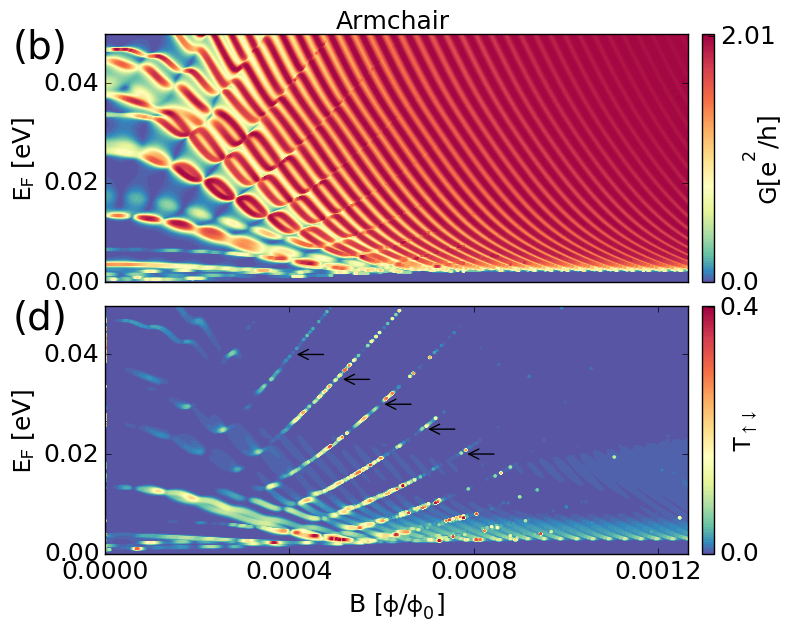}
  \caption{Conductance as a function of magnetic field and Fermi energy in a ring with (a) zigzag and (b) armchair leads, and the spin-flipping component in a ring with (c)  zigzag and (d) armchair leads. The arrows in (c) and (d) indicate the resonances supporting clockwise current in the ring in which the spin flip is enhanced. The inset in (a) shows the schematic spin texture in the ring.
  }  \label{transpBoth}
\end{figure*}

\begin{figure*}[tb!]  
 \includegraphics[width= 0.485\textwidth]{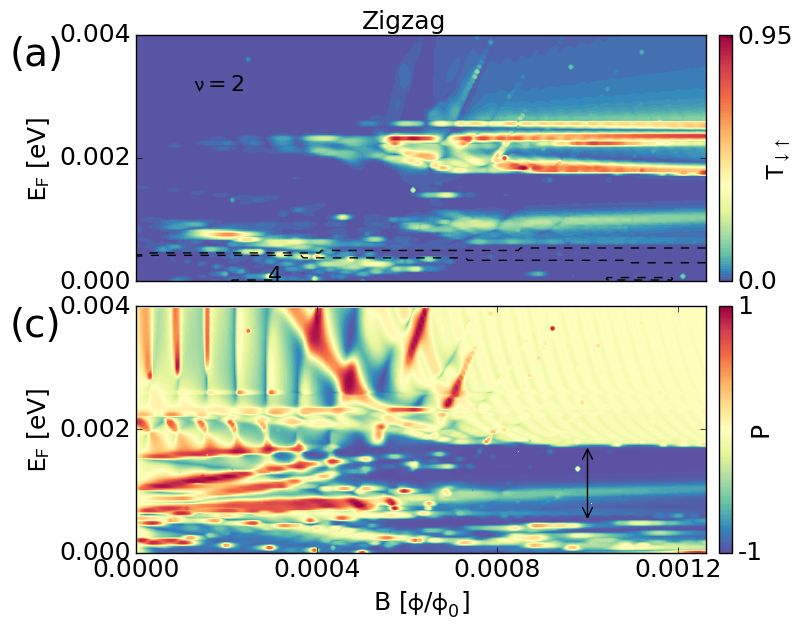}
 \includegraphics[width= 0.49\textwidth]{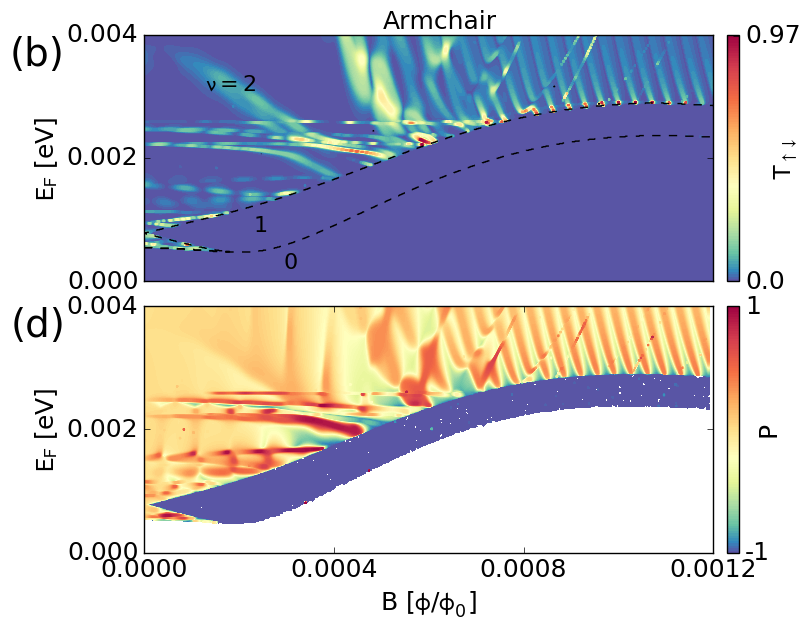}
  \caption{Zoomed (a) and (b) spin-flipping conductance  and (c) and (b) spin polarization as a function of magnetic field and Fermi energy in a ring with zigzag and armchair leads with WSe$_2$. The dashed black lines in (a) and (b) indicate where the number of subbands in the lead changes. The white area in (d) occurs for $G=0$ in the denominator of $P$ in Eq.~(\ref{eq:polarization}).
  }  \label{transpRibbonZoom}
\end{figure*}

Figure \ref{transpBoth} shows the summed conductance [Figs.~\ref{transpBoth}(a) and \ref{transpBoth}(b)] and the spin-flipping conductance [Figs.~\ref{transpBoth}(c) and \ref{transpBoth}(d)] as a function of magnetic field and Fermi energy for the system with zigzag and armchair leads. %In both systems the conductance exhibits two main types of resonances: the ones with an energy growing (decreasing) with magnetic field that correspond to the states with the current flowing near to the outer (inner) edge of the ring \cite{Bahamon2009,Poniedzialek,MrencaLorentz}. 
%In Fig.~\ref{transpBoth}(c,d) the transition between the modes of opposite spin is shown. 
The spin inversion is highest close to the Dirac point and corresponds to the ring eigenstates that have spin aligned in the graphene plane [see Fig.~\ref{eigenringLow}(b)]. This is most pronounced in the system with zigzag leads in which the transport gap is smaller and the transport is mediated by the lowest-lying states with spin almost entirely in the plane [see Fig.~\ref{transpRibbonZoom}(a) for the low-energy zoom]. The incoming states have out-of-plane spin and can flip via those states with the in-plane spin. 
The spin inversion probability for higher energy is generally low [see Fig.~\ref{transpBoth}(c) and \ref{transpBoth}(d)]. 
The exception is the narrow resonances with energy increasing with magnetic field, indicated by black arrows in Figs.~\ref{transpBoth}(c) and \ref{transpBoth}(d). These resonances correspond to the quantum ring states with current circulating around the inner edge of the ring \cite{MrencaLorentz}. Such states have a long lifetime, and the electrons remain a long time in the system, taking many turns around the ring \cite{MrencaLorentz}. The SO in-plane effective magnetic field for the long-living resonances eventually leads to the  spin flips of the Fermi level electron. 
On the other hand, the resonances that circulate close to the outer edge of the ring have a short lifetime because the magnetic field steers the current out of the ring to the right lead. The cumulated phase is not large enough for the spin flip to occur.

% \textcolor{red}{so the spin precession due to the in-plane component of the field  allows for the transition between modes of different spin.} %The spin flip probability is higher in the system with armchair leads because the incoming modes are polarized in the graphene plane, and are non-orthogonal to the states in the ring.

%\textcolor{red}{At low magnetic field the precession axis of the spin is non parallel to the spin -- it has an in-plane component due to the Rashba and PIA SOC, whereas the spin incoming from the zigzag leads is oriented almost entirely in the $z$ direction, and from armchair leads in the $x$ direction. Therefore a finite but small spin rotation occurs. For high magnetic field the rotation axis is almost completely oriented in the $z$ direction, and the spin for zigzag leads is almost parallel to the precession axis. - ale wyjasnienie z precesja nie zgadza sie dla armchaira. Jakos inaczej.}

In addition to the spin flip, we expect that the system can be used for spin filtering. 
%\subsubsection{ Spin polarization for armchair and zigzag edge }
In Fig.~\ref{transpRibbonZoom}(c) the low-energy zoom of the spin polarization in a quantum ring with zigzag leads is shown. At energy below approximately $1.7$meV the modes carrying spin-up states are almost entirely blocked. This is the result of the lack, in this low energy range, of spin-up states supporting clockwise or anticlockwise current. The system exhibits a range of magnetic field in which the current is spin polarized. The energy window of $P\approx -1$ is indicated by the black arrow in Fig.~\ref{transpRibbonZoom}(c). 

%The bound states in the closed ring which maintain clockwise or anticlockwise current, have spin in $z$ direction, unless they are nearly degenerate. Those of these states that have spin-up polarization, disappear for energy below $1.7$meV (see Fig.~\ref{eigenringLow}). As a result, in this energy range the spin-up states almost do not pass through the ring. 

On the other hand, for armchair leads the filtering is more challenging because the armchair lead acquires an energy gap that at high magnetic field saturates at about 3 meV [see Figs.~\ref{transpRibbonZoom}(b) and \ref{transpRibbonZoom}(d)], already above the onset of the ring spin-up levels that carry current around the ring. %Furthermore, the states incoming from the armchair lead are polarized in the ribbon plane, and the spin part of their wave function is not orthogonal to the one in the ring. 
In Fig.~\ref{transpRibbonZoom}(d) the low-energy zoom of the spin polarization in a quantum ring with armchair leads is shown.
In high magnetic field, %only at the onset of the subbands, the spin direction has significant $z$ component [see Fig.~\ref{dispRibbonArm}(d)] but 
the spin subband splitting is only about 0.7 meV. The spin polarization is shown with only a narrow range of $P\approx -1$. The system with armchair leads would require a more accurate tuning of the parameters for the spin filtering.

\begin{figure*}[tb!]
 \includegraphics[width= .49\textwidth]{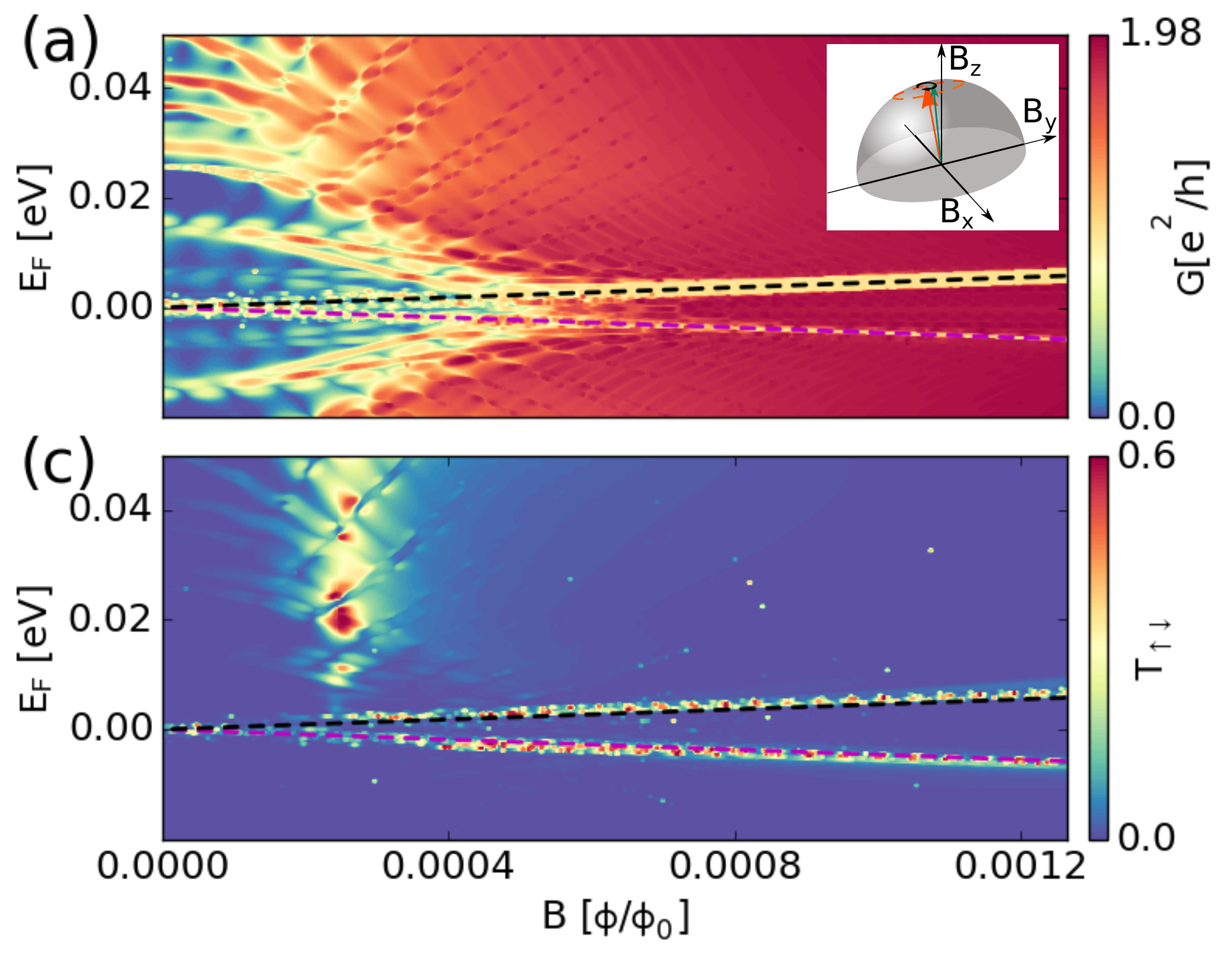}
 \includegraphics[width= .49\textwidth]{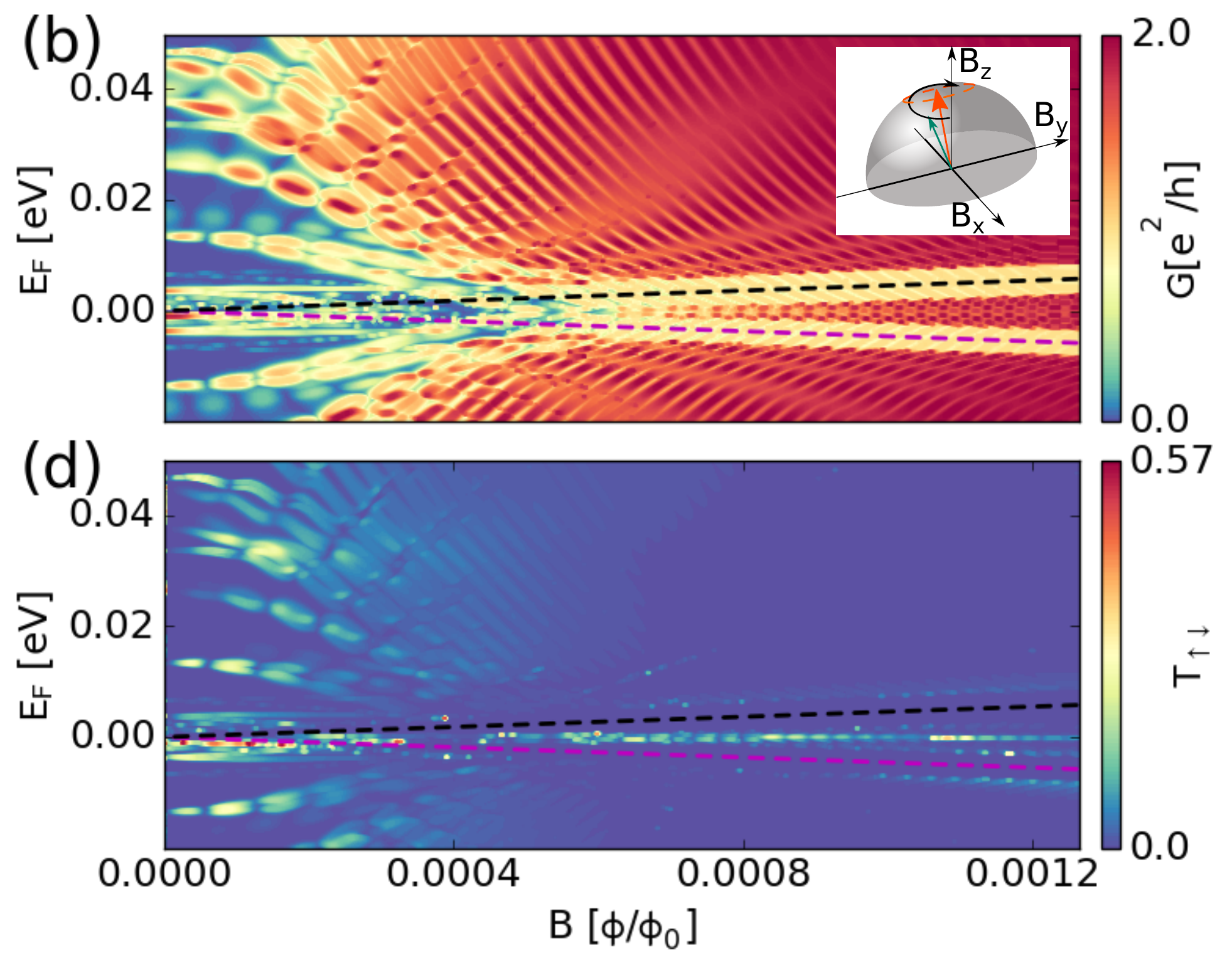}
  \caption{Conductance as a function of magnetic field and Fermi energy in ring with (a) zigzag and (b) armchair leads with Zeeman splitting included. The black (red) dashed lines show the Dirac point for the spin-up (spin-down) electrons. The insets in (a) and (b) show the scheme of the spin precession around the effective magnetic field. The green arrow represents the spin, and the orange arrow indicates the precession axis, which in the rest frame of the electron traverses a circular path around the ring. The black circle with an arrowhead is the momentary precession. The angles between the arrows are exaggerated.
  }  \label{transpBothZeem}
\end{figure*}

In Figs.~\ref{transpRibbonZoom}(a) and \ref{transpRibbonZoom}(b) the zoomed spin-flip probability is shown. 
The transition probability between modes with opposite spin directions is highest close to the narrow resonant states, especially for the zigzag leads [Fig.~\ref{transpRibbonZoom}(a)]; however, obtaining such a high inversion probability would require fine tuning of the back-gate potential and magnetic field. On the other hand, the energy range where the spin filtering occurs is much broader. Therefore we conclude that the quantum rings are more suitable for spin filtering than for spin inversion.

\begin{figure}[tb!]
 \includegraphics[width=0.49\columnwidth]{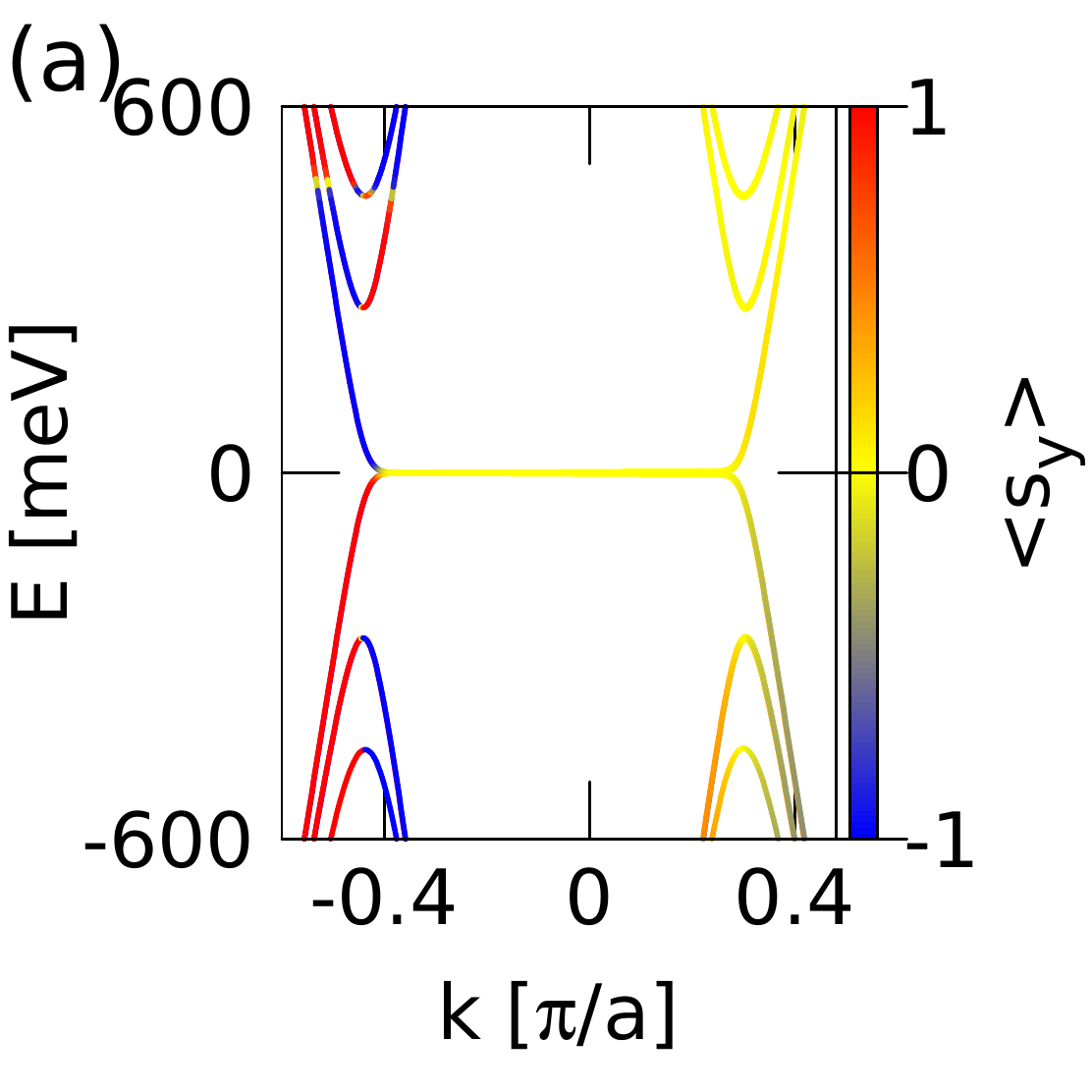}
 \includegraphics[width=0.415 \columnwidth]{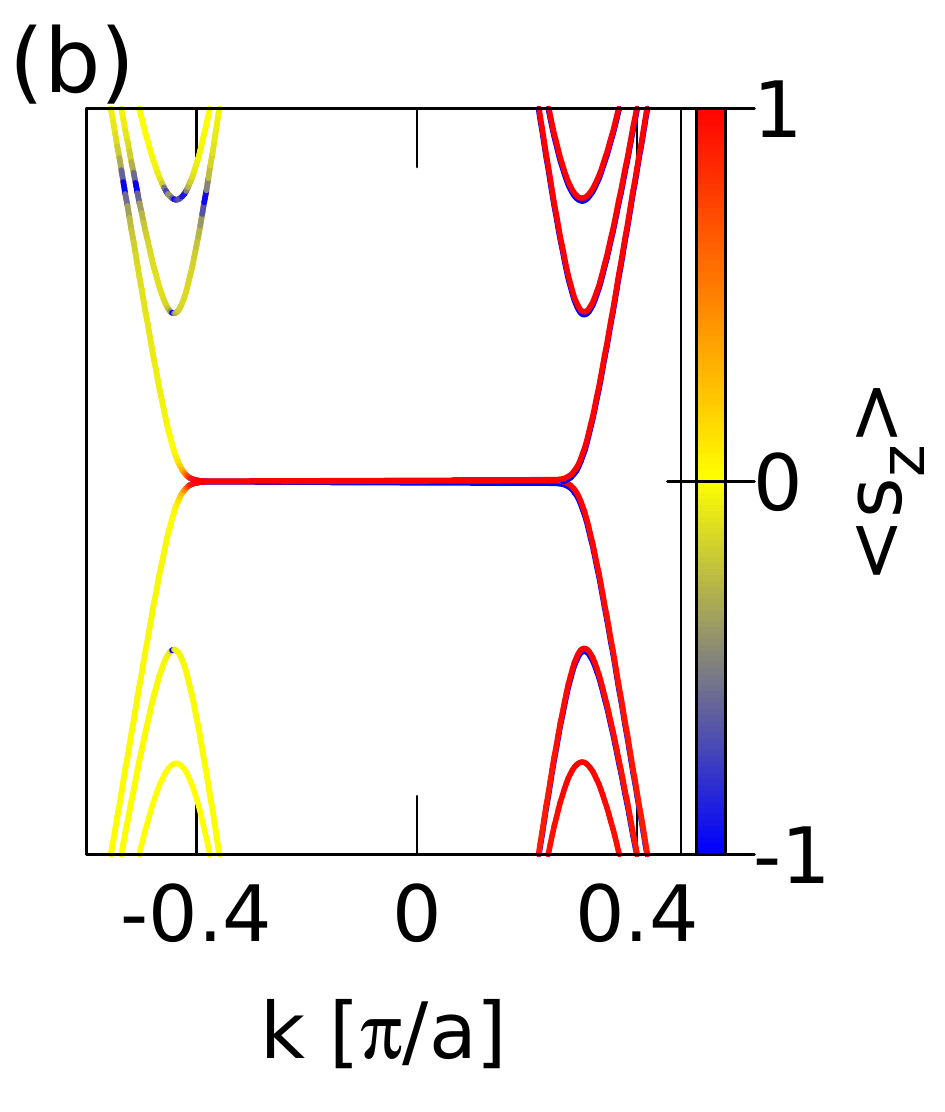}
%\end{center}
  \caption{The band structure of zigzag nanoribbon with WSe$_2$ for $\phi=0.00025 \phi_0$. The color scale shows the mean (a) spin $y$ component and (b) spin $z$ component.
  } \label{zigZeemRel}
\end{figure}

\subsubsection{Transport with the Zeeman effect}
So far we have considered the transport without Zeeman splitting in order to understand the pure effect of SOC.
For completeness, we study the influence of the Zeeman effect on the transport properties of the considered systems. For magnetic field used in experiment, in the range of a few teslas, the Zeeman splitting is of the order $2\times \tfrac{1}{2}g\mu_B B=0.06-0.6$ meV for $B=1-10$ T, with $\mu_B$ being the Bohr magneton, and $g=2$. This is small compared to the maximum SOC-induced splitting of around $2.4$ meV for graphene contacted with WSe$_2$, but both effects add up, and the splitting can reach, for example, 3 meV for $B=10$ T. The Zeeman splitting influences the spin inversion characteristics, as we present below.
%is orders of magnitude higher,and predominates over the SOC-induced splitting.% -- which is almost constant with respect to magnetic field. 

%We discuss the magnetotransport in a system with semiconducting armchair leads 8.61 nm wide (with 142 atoms across the ribbon).
Figure \ref{transpBothZeem} shows the summed conductance and the spin-flipping components of conductance as a function of the Fermi energy of the incident electrons and the external magnetic field. 
%With the Zeeman splitting the spin inversion characteristics change. Firstly, 
In the system with zigzag leads a vertical strip of higher spin flip emerges [Fig.~\ref{transpBothZeem}(c)]. This is when the Zeeman energy coincides with the opposite-spin subbands and the subbands have spin in the plane (see Fig.~\ref{zigZeemRel}) around the value $\phi= 0.00025 \phi_0$. %(20 T). 
Second, the spin flip in the clockwise-current resonances disappear because the Zeeman splitting separates the resonant states for both spins. Without the overlap of these states no spin transfer can occur. Along the ring an effective SO field occurs, shown schematically in the inset of Fig.~\ref{transpBoth}(a). In high magnetic field the spin of the incoming electrons from both types of leads is oriented more in the $z$ direction, close to the precession axis [see the inset of Figs.~\ref{transpBothZeem}(a) and \ref{transpBothZeem}(b)]; therefore no spin-flip occurs.
For zigzag ribbons [Figs.~\ref{transpBothZeem}(a) and \ref{transpBothZeem}(c)] the effective  magnetic field due to the SOC superposes with the Zeeman effect. For the armchair ribbons the incident spins 
at low magnetic field are deflected to the in-plane orientation by the Rashba and PIA SOC [see the inset of Fig.~\ref{transpBothZeem}(b)], and the variation of the spin within the ring appears via precession 
in the effective intrinsic SO magnetic field oriented in the $z$ direction, which is missing in the leads due to the intervalley scattering. At higher magnetic field the spin-flipping transport disappears
when the Zeeman interaction dominates over the effective SO interaction.

%\textcolor{red}{To do ewentualnie supplementu?: Fig.~\ref{transpZeemZig} show the low energy zoom in the system with zigzag leads. 
% In Fig.~\ref{transpZeemZig}(a,c), the spin-conserving components are shown with for spin down [Fig.~\ref{transpZeemZig}(a)] or spin up [Fig.~\ref{transpZeemZig}(c)] component. Note that there is a transport gap that is different for both spin directions. The gap is due to the spectrum of the closed ring. Without Zeeman splitting, the ring states with smallest energy $|E|$ originate from $K$ ($K'$) valley for $E<0$ ($E>0$), having opposite direction of the SOC field [Eq.~(\ref{eq:SOfield})]. For spin up the energy levels above Dirac point have higher energy, and below the Dirac point -- lower energy, and for spin down this is reversed. Thus for spin up the gap is bigger than for spin down. With Zeeman interaction included, the energy of the spin-up (spin-down) states increases (decreases) with magnetic field. Thus the gap around the spin-up Dirac point is bigger. 
% The low-energy spin-flipping resonances are visible in Fig.~\ref{transpZeemZig}(b) but they are doubled due to the Zeeman splitting. In Fig.~\ref{transpZeemZig}(d) the polarization is shown with two stripes of spin filtering action for spin up and down electrons clearly visible. }

\section{Summary and Conclusions}

We considered the application of a graphene-TMDC heterostructure for building spin-active elements. We studied the properties of quantum rings produced by graphene in contact with WSe$_2$. The induced valley Zeeman SO coupling leads to energy splitting of the ring levels of opposite spin. In magnetic field the system has spin-filtering properties when tuned to the Fermi energy between the split levels. For this purpose the zigzag leads are especially promising because the zigzag edge does not mix the $K$ and $K'$ valleys. Quantum rings can also be used as a spin-inverting element for building a spin transistor; however, for the complete spin inversion high precision of the electron energy or external magnetic field would be required.

\section*{Acknowledgments}
This work was supported by the National Science Centre (NCN) according to decision DEC-2015/17/B/ST3/01161
and by AGH UST budget with the subsidy of the Ministry of
Science and Higher Education, Poland with Grant No. 15.11.220.718/6 for young researchers 
and Statutory Task No. 11.11.220.01/2.
The calculations were performed on PL-Grid Infrastructure.

\bibliographystyle{apsrev4-1}
\bibliography{RingTMDC}

\end{document}